# A Method for Predicting VaR by Aggregating Generalized Distributions Driven by the Dynamic Conditional Score

Shijia Song
202131250022@mail.bnu.edu.cn
School of Systems Science, Beijing Normal University, Beijing 100875, China

Handong Li
lhd@bnu.edu.cn
School of Systems Science, Beijing Normal University, Beijing 100875, China

Correspondence should be addressed to Handong Li
E-mail: lhd@bnu.edu.cn
Tel. 8610-58807064
Tax. 8610-58807876

# A Method for Predicting VaR by Aggregating Generalized Distributions Driven by the Dynamic Conditional Score

**Abstract:**
Constructing a more effective value at risk (VaR) prediction model has long been a goal in financial risk management. In this paper, we propose a novel parametric approach and provide a standard paradigm to demonstrate the modeling. We establish a dynamic conditional score (DCS) model based on high-frequency data and a generalized distribution (GD), namely, the GD-DCS model, to improve the forecasts of daily VaR. The model assumes that intraday returns at different moments are independent of each other and obey the same kind of GD, whose dynamic parameters are driven by DCS. By predicting the motion law of the time-varying parameters, the conditional distribution of intraday returns is determined; then, the bootstrap method is used to simulate daily returns. An empirical analysis using data from the Chinese stock market shows that Weibull-Pareto -DCS model incorporating high-frequency data is superior to traditional benchmark models, such as RGARCH, in the prediction of VaR at high risk levels, which proves that this approach contributes to the improvement of risk measurement tools.

## 1. Introduction

The wide application of electronic trading systems in financial markets and the massive increase in the quantity of business trading have made market risk measurement the main focus of regulatory authorities. Risk measurement can provide banks and financial institutions with specific potential loss values so that risk managers can adjust capital reserves for downside risks. VaR is one of the most important financial risk measurement tools that dominate contemporary financial supervision. VaR provides the worst-case loss at a given level of confidence. According to the Basel Accord proposed by the Bank for International Settlements (BIS) in 1996, a bank's risk capital must be sufficient to cover 99% of the possible losses during a 10-day holding period. VaR has become the most popular risk management tool in the financial services industry; however, VaR is inherently flawed because it ignores the shape and structure of the tail. Artzner et al. (1999) believed that VaR was not a coherent risk measurement and that it cannot accurately measure market risk. Although other risk measures, such as expected shortfall (ES), can make up for the shortcomings of VaR to some degree, they cannot completely replace VaR and shake the position of VaR as one of the most important risk management tools. Therefore, improving the accuracy of VaR forecasts remains a core issue in financial risk measurement.

Combined with the expression of Engle and Manganelli (2004), the current models of risk measurement can be roughly divided into three categories: parametric, nonparametric and semiparametric models. Since the returns of financial market variables usually exhibit nonnormality, parametric models are often criticized for failing to specify the correct distributions for these variables. Nonparametric models mostly construct portfolio models based on historical returns of a specific window length to mimic the past performance of current portfolios and then calculate current VaR based on statistical models. Such models do not require distributional assumptions, but the best size for the estimation window is difficult to determine (Engle and Manganelli, 2004). Some recent semiparametric models directly impose a dynamic parameter structure on VaR without assuming the conditional distribution of financial returns (Engle and Manganelli, 2004; Patton et al., 2019). Based on the semiparametric model proposed by Patton et al. (2019), Lazar and Xue (2020) added realized volatility to the model for a more accurate joint estimation of VaR and ES. However, due to the high integration of semiparametric models and the limitations of nonparametric models, if we consider further improving the model by directly using intraday high-frequency returns instead of volatility, parametric models are the most feasible.

Whether high-frequency (HF) intraday information can improve the accuracy of risk measurement forecasts has been widely discussed. Since Andersen and Bollerslev (1997), Andersen et al. (2001a), Andersen et al. (2001b), and Barndorff-Nielsen and Shephard (2002) have all used realized volatility as an effective and consistent estimator of potential volatility, the availability of HF data has gradually improved. Shephard and Sheppard (2010), Noureldin et al. (2012), and Hansen et al. (2014) proposed models that include HF to fit the conditional second moment of returns. Bee et al. (2019) added several realized volatility measures to POT models and compared their estimation efficiency for the tail parameter. However, the direct utilization of intraday HF returns in the risk model is still in its infancy. Hallam and Olmo (2014) and Cai et al. (2019) constructed functional autoregressive VaR models to fit the distribution of daily returns by estimating the kernel density function of intraday returns. This background prompts the research question of this article, namely, whether a parametric risk model based directly on intraday returns will improve the accuracy of VaR forecasts. The focus of modeling in our work is then shifted to three specific questions: (a). Which kind of distribution is most suitable for fitting the returns? (b). What mechanism can be used to accurately estimate the time-varying parameters. (c). What method can be used to improve the daily VaR forecast directly using intraday HF returns rather than realized volatility?

Which kind of distributions should be specified to model returns is indeed a question worth considering. Since many of the most important differences between actual distributions and normal distributions are reflected in the "skewness", the early expansion of the distribution is reflected in the construction of the so-called "skew probability curve" system (Pearson, 1895; Burr, 1942; Johnson, 1949). Alzaatreh (2013) introduced a more general method to derive the generalized distribution (GD) family using a probability density function of a random variable with any values. All the GDs obtained through this transformation are called "T-X" distributions. The "T-X" transformation makes it possible to continuously develop newly GDs. This kind of distribution may be flexible and applicable for specific types of data, such as data with a bimodal distribution and financial returns with severe heavy tails.

Although many well-known GDs exist, due to the differences in their generation mechanisms, we consider several "T-X" distributions with the same "X" but different "T" as the distribution pool for fitting financial returns to illustrate the modeling process and compare the performance of different models. The Pareto distribution, a type of statistical model with a power-law tail, is often used to simulate data with obvious right-skewness and a heavy right tail (Klugman, 1998). In the Pareto family, Pareto IV (Cronin, 1979) is particularly worthy of attention. Pareto IV contains the most parameters, and when the parameters in Pareto IV take specific values, the corresponding special cases are named after Pareto I-III. Since Pareto IV can accurately capture the tail risk of financial data, we consider it as the fixed "X" in the various "T-X" distributions to achieve an accurate fitting of the returns.

In financial risk management, it is essential to accurately estimate the time-varying parameters that control the appearance of the returns' distribution. Cox (1981) divided models with dynamic parameters into parameter-driven and observation-driven models. The evolution of parameters in the observation-driven model depends on the function of the observed values, such as the autoregressive conditional heteroscedasticity (ARCH) model of Engle (1982), the generalized autoregressive conditional heteroscedasticity (GARCH) of Bollerslev (1986) and the dynamic conditional scoring model (DCS) proposed by Creal et al. (2012) and Harvey (2013). The factor that drives the parameters in the DCS model is the standardized score (namely, the first derivative of the probability density function with respect to the parameter), which ensures strong adaptability to non-normal data and maintains highly robust estimation (Lucas et al., 2015). Thus, DCS has been increasingly popular in fitting the distribution of financial variables. Zhang and Bernd (2016) and Massacci (2017) introduced dynamic score-driven models to estimate the probability of extreme returns and the size of the exceedance. Ayala et al. (2019) proposed a DCS model based on the normal inverse Gaussian (NIG) distribution that can simultaneously update the volatility through the scale parameter and the shape parameter. Patton et al. (2019) developed methods for the joint assessment of dynamic VaR and ES under the framework of DCS. Encouraged by the fact that the DCS model is suitable for fitting distributions with time-varying parameters and has highly efficient estimation, we consider constructing a model with DCS to better capture the motion law of returns.

Most applications of HF information in risk modeling consider realized volatility based on intraday returns as an extra explanatory variable or make use of the deviation of HF data to improve the semiparametric risk model. Under the framework for the joint estimation of VaR and ES proposed by Patton et al. (2019), Lazar and Xue (2020) added realized volatility to the original quantile regression setup to estimate risk measures. Rice et al. (2020) focused on the deviation of intraday returns and constructed a VaR model based on GARCH and a functional quantile regression model. Inspired by the research of Cai et al. (2019), instead of converting intraday information into volatility, we apply bootstrapping to directly sample intraday returns to obtain the simulated distribution of daily returns rather than apply the common method. Then, VaR is obtained through the quantiles of the distribution. Several backtesting procedures and model confidence sets (MCSs) are implemented to determine the relatively best frequency of intraday returns to be used in modeling and to compare the forecasting effects of GD-DCS-VaR and RGARCH-VaR.

The remainder of this paper is organized as follows. Section 2 outlines the construction and assessment of the GD-DCS model. It first introduces three "T-Pareto IV" generalized distributions and gives the score-driven equation for each dynamic parameter in these GDs along with its maximum likelihood (ML) estimator. Then, it explains the bootstrap method for obtaining the daily return distribution based on intraday returns. Finally, it illustrates three common backtesting approaches and MCS for assessing the effect of out-of-sample VaR forecasts. Section 3 details the empirical data of the Chinese stock market used in the analysis and the corresponding data processing. In Section 4, the test results indicate that VaR forecasts by the model formed on HF data are indeed less likely to underestimate risk, and in terms of coverage ability and MCS, GD-DCS-VaR gains an advantage over RGARCH-VaR at high risk levels. Section 5 concludes this article. Supplementary materials are relegated to Appendices A and B.

## 2. Methodology

### 2.1 "T-X" family

Alzaatrech (2013) proposed a method for generating continuous GD families that allows the probability density function (p.d.f.) of any distribution to be used as a generator.

Denote $r(t)$ as the p.d.f. of a random variable $T$, $T \in [a, b]$, and denote $W(F(x))$ as the cumulative distribution function (c.d.f.) of any continuous random variable $X$. $W(F(x))$ should satisfy $W(F(x)) \in [a, b]$; $W(F(x))$ is differentiable, monotonous and nondecreasing; $\lim_{x \to -\infty} W(F(x)) = a$, $\lim_{x \to \infty} W(F(x)) = b$. Then, the c.d.f. of a new GD family can be deduced by the following formula:

$$G(x) = \int_a^{W(F(x))} r(t)\, dt \tag{1}$$

Correspondingly, the p.d.f. is

$$g(x) = \left\{\frac{d}{dx} W(F(x))\right\} r\{W(F(x))\} \tag{2}$$

$W(F(x))$ acts as a "transformer" to create the p.d.f. $r(t)$ is "transformed" into a new c.d.f., $G(x)$, through integration in (1). Therefore, $g(x)$ in (2) has completed the transformation from the random variable $T$ to random variable $X$, and the GD defined by (1) is named the "Transformed-Transformer" or "$T - X$" distribution.

Different $W(F(x))$ can define different GDs. The specific form depends on the value range of the random variable $T$; for details, please refer to the definition of Alzaatrech (2013). Since common distributions used as the "transformed" distribution, such as the Weibull distribution, gamma distribution, and Rayleigh distribution, all require nonnegative observations, for $T \in [0, +\infty)$, $W(F(x))$ has three commonly used expressions, among which $W(F(x)) = -\ln(1 - F(x))$ is the form focused on by Alzaatrech (2013). This paper also focuses on the GD family derived from $T \in [0, +\infty)$ and $W(F(x)) = -\ln(1 - F(x))$. In addition, to make the different GDs in the GD pool more comparable, we fix X in T-X, that is, we let X be a random variable that obeys the Pareto IV distribution, and then compare the fitting efficiency with different T's.

Different $W(F(x))$ can define different GDs, and its specific form depends on the range of the random variable $T$. More details can be found in the definition of Alzaatrech (2013). For example, for distributions that are commonly "transformed", such as the Weibull distribution, gamma distribution, and Rayleigh distribution, since they require nonnegative observations, i.e., $T \in [0, +\infty)$, $W(F(x))$ has three corresponding expressions. Among them, $W(F(x)) = -\ln(1 - F(x))$ is the major form that Alzaatrech (2013) focused on and is also the form we are interested in. Specifically, to make the different GDs in the "pool" more comparable, we use three different distributions with a fixed "$X$"; that is, we let $X$ be a random variable that obeys the Pareto IV distribution and then compare the fitting efficiency of T-Pareto IV GDs, where "$T$" refers to the Weibull, gamma, and Rayleigh distributions, respectively.

### 2.2.1 Weibull-Pareto IV

Assuming that the random variable $T$ follows a Weibull distribution, its p.d.f. is

$$r(t) = \frac{c}{\gamma}\left(\frac{t}{\gamma}\right)^{c-1} \exp\left\{-\left(\frac{t}{\gamma}\right)^c\right\}, t \geq 0, c \geq 0, \gamma \geq 0 \tag{3}$$

Since

$$G(x) = \int_0^{-\ln(1-F(x))} r(t)dt \tag{4}$$

Hence, the p.d.f. and c.d.f. of the Weibull-Pareto IV can be derived as:

$$G(x) = 1 - \exp\left\{-\left[\frac{-\ln(1-F(x))}{\gamma}\right]^c\right\} \tag{5}$$

$$g(x) = \frac{c}{\gamma}\frac{f(x)}{1-F(x)}\left[\frac{-\ln(1-F(x))}{\gamma}\right]^{c-1}\cdot\exp\left\{-\left[\frac{-\ln(1-F(x))}{\gamma}\right]^c\right\} \tag{6}$$

If $X$ obeys Pareto IV, $f(x) = \frac{\delta}{\alpha}(x)^{\frac{1}{\alpha}-1}[1+x^{\frac{1}{\alpha}}]^{-\delta-1}, x > 0, \delta > 0, \alpha > 0$; then

$$g(x) = \frac{c}{\alpha}\left(\frac{\delta}{\gamma}\right)^c x^{\frac{1}{\alpha}-1}\left(1+x^{\frac{1}{\alpha}}\right)^{-1}\left[\ln\left(1+x^{\frac{1}{\alpha}}\right)\right]^{c-1}\cdot\exp\left\{-\left[\frac{\delta}{\gamma}\ln\left(1+x^{\frac{1}{\alpha}}\right)\right]^c\right\} \tag{7}$$

Let $\beta = \frac{\delta}{\gamma}$; equation (7) can then be expressed as:

$$g(x) = \frac{c\beta^c}{\alpha}x^{\frac{1}{\alpha}-1}\left(1+x^{\frac{1}{\alpha}}\right)^{-1}\left[\ln\left(1+x^{\frac{1}{\alpha}}\right)\right]^{c-1}\cdot\exp\left\{-\left[\beta\ln\left(1+x^{\frac{1}{\alpha}}\right)\right]^c\right\} \quad x > 0,\ \alpha,\beta,c > 0 \tag{8}$$

The distribution that satisfies the above formula is the so-called Weibull-Pareto IV, denoted as $WPD^{IV}(\beta,\alpha,c)$, where $\beta$ is the scale parameter, $c$ is the shape parameter, and $\alpha$ is called the inequality parameter because of its interpretation in the economics context. The c.d.f. of $WPD^{IV}$ is:

$$G(x) = 1 - \exp\left\{-\left[\beta\ln\left(1+x^{\frac{1}{\alpha}}\right)\right]^c\right\} \quad x > 0,\ \alpha,\beta,c > 0 \tag{9}$$

**2.1.2 Gamma-Pareto IV**

Assuming that $T \sim Gamma(\theta,\beta)$, its p.d.f. is

$$r(t) = (\beta^\theta \Gamma(\theta))^{-1} t^{\theta-1} e^{-\frac{t}{\beta}}, t \geq 0, \theta > 0, \beta > 0 \tag{10}$$

In terms of equation (4), the p.d.f. of Gamma-X is:

$$g(x) = \frac{f(x)}{1-F(x)} r\left(-\ln(1-F(x))\right) = \frac{f(x)(-\ln(1-F(x)))^{\theta-1}(1-F(x))^{\frac{1}{\beta}-1}}{\Gamma(\theta)\beta^\theta} \tag{11}$$

Similarly, if $X \sim$ Pareto IV$(\delta,\alpha)$, the p.d.f. of Gamma-Pareto IV is:

$$g(x) = \frac{1}{\alpha c^\theta \Gamma(\theta)} (x)^{\frac{1}{\alpha}-1} (1+x^{\frac{1}{\alpha}})^{-1-\frac{1}{c}} \left[ \ln\left(1+x^{\frac{1}{\alpha}}\right) \right]^{\theta-1}, x > 0, \alpha > 0, c > 0, \theta > 0, \quad (12)$$

where $c = \frac{\beta}{\delta}$. Distributions with the above p.d.f. are Gamma-Pareto IV distributions, denoted as $GPD^{IV}(\theta, \alpha, c)$, where $\theta, \alpha$ and $c$ denote the scale parameter, inequality parameter, and shape parameter, respectively. The c.d.f. is given by:

$$G(x) = \frac{\gamma\left\{\theta, c^{-1} \ln\left(1+x^{\frac{1}{\alpha}}\right)\right\}}{\Gamma(\theta)}, \quad (13)$$

where function $\gamma(\theta, t) = \int_0^t u^{\theta-1} e^{-u} du$ denotes the incomplete gamma function.

### 2.1.3 Rayleigh-Pareto IV

If $T$ follows a Rayleigh distribution, its p.d.f. is given by $r(t) = \frac{t}{\sigma^2} e^{-\frac{t^2}{2\sigma^2}}, t \geq 0$. From (4), it follows that the p.d.f. of Rayleigh-X is:

$$g(x) = \frac{f(x)}{1-F(x)} r\left(-\ln(1-F(x))\right) = \frac{-f(x)\ln(1-F(x))}{\sigma^2(1-F(x))} \exp\left(\frac{[\ln(1-F(x))]^2}{2\sigma^2}\right) \quad (14)$$

With $X \sim$ Pareto IV$(\delta, \alpha)$, the p.d.f. of Rayleigh-Pareto IV becomes:

$$g(x) = \frac{\delta^2 \ln(1+x^{\frac{1}{\alpha}})}{\sigma^2 \alpha} (x)^{\frac{1}{\alpha}-1} (1+x^{\frac{1}{\alpha}})^{-1} \exp\left(-\frac{\delta^2 \left[\ln\left(1+x^{\frac{1}{\alpha}}\right)\right]^2}{2\sigma^2}\right) x > 0, \alpha > 0, \delta > 0 \quad (15)$$

The distribution defined by (15) is called Rayleigh-Pareto IV, denoted as $RPD^{IV}(\sigma, \alpha, \delta)$, where $\sigma, \alpha$ and $\delta$ denote the scale parameter, inequality parameter, and shape parameter, respectively. The c.d.f. is then given by:

$$G(x) = 1 - \exp\left(-\frac{\delta^2 \left[\ln\left(1+x^{\frac{1}{\alpha}}\right)\right]^2}{2\sigma^2}\right) x > 0, \alpha > 0, \delta > 0 \quad (16)$$

### 2.2 GD-DCS model

We intend to establish a DCS model to estimate the main time-varying parameters appearing in the GDs. DCS is of great interest because the score provides a natural update mechanism that links the dynamics of the parameters with the likelihood of the observed samples (Creal et al., 2012). In risk management, it is essential to grasp the law of motion of parameters that control the pattern of returns. However, there is no clear standard for setting a certain parameter as static or dynamic, which should depend on the specific situation. Some financial models (Engle, 1982; Bollerslev, 1986; Nelson, 1991; Harvey and Shephard, 1996; Barndorff-Nielsen and Shephard, 2002; Bee et al. 2019) assume a dynamic scale parameter but a static shape parameter. Others (Massacci,

2017; Ayala et al., 2019; Harvey and Ito, 2020) set both the scale parameter and shape parameter to be dynamic and achieve better estimation results. The empirical results in this article also proves the effectiveness of considering time-varying scale and shape parameters.

Let $R_{\tau,t}$ denote the $\tau^{th}$ observation of intraday returns on the $t$th day, where $t = 1, \ldots, T$, $\tau = 1, \ldots, N$, and N varies due to the frequency of intraday returns. Considering that the GD mentioned in the article is applicable to variables whose observations are greater than 0, we shift the entire series of observations of $R_{\tau,t}$ a certain number of units to the right to make all the values greater than 0 and then shift the VaR calculated based on the distribution after translation to the left. From (8), if $X_{\tau,t} \sim WPD^{IV}(\alpha, \beta, c)$, the logarithmic expression of its p.d.f. yields:

$$\ln g(x) = \ln c + c \ln \beta - \ln \alpha + \left(\frac{1}{\alpha} - 1\right) \ln x + (c-1) \ln \left[\ln \left(1 + x^{\frac{1}{\alpha}}\right)\right] - \left[\beta \ln \left(1 + x^{\frac{1}{\alpha}}\right)\right]^c \quad x, \alpha, \beta, c > 0 \quad (17)$$

To ensure that both the scale parameter $\beta$ and the shape parameter $c$ are positive, we set $\beta = \exp(\lambda)$, $c = \exp(v)$. Since there may exist an annual cycle structure in returns over a long period of time, to reduce the possible impact that the correlation of the high-frequency return series may bring about, we refer to the practice of Harvey and Ito (2013) and consider adding seasonal factor $q_t$ to the autoregressive equation of the scale parameter. The deterministic $q_t$ can be easily obtained by decomposing the time series. Then, the laws of motion for $\beta_t$ and $c_t$ are specified in terms of an autoregressive process in exponential form as:

$$\begin{cases} \ln \beta_t = \lambda_t = A_1 + B_1 \lambda_{t-1} + C_1 s_{\lambda,t} + q_t \\ \ln c_t = v_t = A_2 + B_2 v_{t-1} + C_2 s_{v,t} \end{cases}, \quad (18)$$

where $s_{\lambda,t}$ and $s_{v,t}$ refer to the standardized scores of $\lambda_t$ and $v_t$. Then, it follows that:

$$\begin{cases} s_{\lambda,t} = \nabla_{\lambda,t} \cdot S_{\lambda,t} = \nabla_{\beta,t} \cdot \exp(\lambda_t) \cdot \dfrac{1}{E\left(\dfrac{\partial^2 \ln g(x)}{\partial \beta^2}\right) \cdot \exp(2\lambda_t)} \\ s_{v,t} = \nabla_{v,t} \cdot S_{v,t} = \nabla_{c,t} \cdot \exp(v_t) \cdot \dfrac{1}{E\left(\dfrac{\partial^2 \ln g(x)}{\partial c^2}\right) \cdot \exp(2v_t)} \end{cases}, \quad (19)$$

where $\nabla_{\lambda,t}$ and $\nabla_{v,t}$ denote the first-order partial derivatives of the logarithmic density function (17) with respect to the parameters and $S_{\lambda,t}$ and $S_{v,t}$ represent the factors used to standardize the scores.

Similarly, for $X_{\tau,t} \sim GPD^{IV}(\theta, \alpha, c)$, we write its logarithmic density function as:

$$\ln g(x) = -[\ln \alpha + \theta \ln c + \ln \Gamma(\theta)] + \left(\frac{1}{\alpha} - 1\right) \ln x + \left(-1 - \frac{1}{c}\right) \ln \left(1 + x^{\frac{1}{\alpha}}\right) + (\theta - 1) \ln \left[\ln \left(1 + x^{\frac{1}{\alpha}}\right)\right] \quad x, \alpha, c, \theta > 0 \quad (20)$$

Due to the positiveness of $c$ and $\theta$, we let $c = \exp(\lambda)$, $\theta = \exp(v)$ and define their score-driven models as:

$$\begin{cases} \ln c_t = \lambda_t = A_1 + B_1 \lambda_{t-1} + C_1 s_{\lambda,t} \\ \ln \theta_t = v_t = A_2 + B_2 v_{t-1} + C_2 s_{v,t} + q_t \end{cases}, \quad (21)$$

where

$$\begin{cases} s_{\lambda,t} = \nabla_{\lambda,t} \cdot S_{\lambda,t} = \nabla_{c,t} \cdot \exp(\lambda_t) \dfrac{1}{E\left(\frac{\partial^2 \ln g(x)}{\partial c^2}\right)\cdot\exp(2\lambda_t)} \\ s_{v,t} = \nabla_{v,t} \cdot S_{v,t} = \nabla_{\theta,t} \cdot \exp(v_t) \cdot \dfrac{1}{E\left(\frac{\partial^2 \ln g(x)}{\partial \theta^2}\right)\cdot\exp(2v_t)} \end{cases} \quad (22)$$

If $X_{\tau,t} \sim RPD^{IV}(\sigma, \alpha, \delta)$, then the p.d.f. of this random variable can be expressed as follows:

$$\ln g(x) = 2\ln\delta + 2\ln\left[\ln\left(1 + x^{\frac{1}{\alpha}}\right)\right] - 2\ln\sigma - \ln\alpha + \left(\frac{1}{\alpha} - 1\right)\ln x - \ln\left(1 + x^{\frac{1}{\alpha}}\right) - \dfrac{\delta^2\left[\ln\left(1 + x^{\frac{1}{\alpha}}\right)\right]^2}{2\sigma^2} \quad x > 0, \alpha > 0, \delta > 0 \quad (23)$$

Denote $\delta = \exp(v)$ and construct the score-driven model as:

$$\begin{cases} \sigma_t = A_1 + B_1\sigma_{t-1} + C_1 s_{\sigma,t} + q_t \\ \ln\delta_t = v_t = A_2 + B_2 v_{t-1} + C_2 s_{v,t} \end{cases}, \quad (24)$$

where

$$\begin{cases} s_{\sigma,t} = \nabla_{\sigma,t} \cdot S_{\sigma,t} = \nabla_{\sigma,t} \dfrac{1}{E\left(\frac{\partial^2 \ln g(x)}{\partial \sigma^2}\right)} \\ s_{v,t} = \nabla_{v,t} \cdot S_{v,t} = \nabla_{\delta,t} \cdot \exp(v_t) \cdot \dfrac{1}{E\left(\frac{\partial^2 \ln g(x)}{\partial \delta^2}\right)\cdot\exp(2v_t)} \end{cases} \quad (25)$$

See Appendix A for the specific derivation process involved in the formulas. Then, the ML method is used to estimate the combination of parameters, $\emptyset = \{A_1, B_1, C_1, A_2, B_2, C_2\}$, involved in the GD-DCS model, namely:

$$\widehat{\emptyset} = argmax \sum_{t=1}^{T} \ln[f_t(r_t|r_1, \dots, r_{t-1})], t = 1,2,3 \dots T \quad (26)$$

### 2.3 Intraday-return-based Estimation for Daily VaR

The DP-DCS model specifies a GD for intraday returns and estimates the time-varying parameters in each interval and on each day. Since the form of the HF distribution is determined via this approach, how to combine the information of intraday returns to fit the conditional distribution of daily returns deserves careful consideration. Furthermore, which frequency of intraday data provides the most accurate estimate also needs to be verified.

In terms of the strong form of efficiency market hypothesis, intraday returns at different moments can be independent from each other. Under this strong assumption, a common statistical method named bootstrap can be used to simulate a conditional distribution of daily return based on the sum of samples of intraday returns. The essence of this method is to directly add up the moment-specific simulated returns to get the daily return. However, due to the possible correlation between the returns within the same day, there will be overlaps in the aggregation process, which will result in errors in the estimation of the daily return. As mentioned in section 2.2, we believe that the intraday periodic structure may be the cause of the linear correlation between the series, so the seasonal factor is added to the model to relieve the impact of independence and thus make the method more feasible.

The bootstrapping takes the following specific steps: First, the probability density distribution of the $\tau^{th}$ intraday returns is divided into M equal areas; then, M non-equidistant bins are produced and their intervals are calculated to form a set, $\{x_{\tau,0}, \ldots, x_{\tau,j}, \ldots, x_{\tau,M}\}$, $j = 0,1,\ldots,M$. Second, we define $m_{\tau,j} = (x_{\tau,j} + x_{\tau,j+1})/2$ and form a grid set, $\{m_{\tau,1}, \ldots, m_{\tau,j}, \ldots, m_{\tau,M}\}$. Third, we draw N (the number of HF data points within one day) real numbers randomly from a uniform distribution over (0, 1) for each bootstrap iteration; thus, the numbers can be denoted as $\{q_{1,b}, \ldots, q_{\tau,b}, \ldots, q_{N,b}\}$, where $q_{\tau,b} \sim U(0,1)$ and the number of iterations satisfy $b = 1, \ldots, B$. By solving the inverse function of (4) with $q_{\tau,b}$ as the independent random variable, we can obtain an approximate intraday return. Next, the approximate return is compared with $x_{\tau,j}$ and we determine which grid the return belongs to in terms of the corresponding $m_{\tau,j*}$ as the estimated $\tau^{th}$ intraday return. Therefore, a daily return can be constructed by $R_{b,t} = \sum_{\tau=1}^{N} m_{\tau,j*}$ for each iteration. After B bootstrapping cycles, we obtain the estimated distribution of daily returns on day $t$ when B is sufficiently large. Generally, M is set to 100 and B is set to 1000. Finally, we treat the $\alpha$th quantile of this distribution as the estimated daily $\widehat{VaR}(\alpha)$. Similarly, the same approach can be applied to intraday returns on other days to generate a series of $\widehat{VaR}(\alpha)$ values.

The GD-DCS-VaR model is essentially a parametric approach to obtain dynamic daily VaR based on HF returns. Due to the updating mechanism of DCS, the proposed model is expected to be sensitive to variation in risk. The incorporation of intraday information allows it to measure risk from a microscopic perspective. Theoretically, this GD-DCS-VaR model could precisely measure the tail risk of returns and thus improve VaR forecasts.

**Backtesting and MCS for VaR**

Several backtesting methods can be used to assess the estimates or forecasts of VaR generated by the GD-DCS model. Backtesting is a statistical procedure where actual profits and losses are systematically compared to the corresponding VaR estimates. We consider applying the LR of the unconditional coverage test (LRUC), the LR of the conditional coverage test (LRCC), and the dynamic quantile (DQ) test to measure the coverage ability and the independence of VaR.

However, backtesting fails to provide an overall assessment of the effectiveness of VaR; thus, it is difficult to intuitively compare the performance of different models. Hansen et al. (2011) developed the MCS procedure to assess the performance of a given set of VaR series belonging to several different models. Hansen's procedure can yield a set of "superior" models based on a series of statistical tests, where the null hypothesis of equal predictive ability (EPA) is not rejected at a certain confidence level. We could then perform MCS to comprehensively evaluate the accuracy and effectiveness of several VaR models.

## 3. Data and data processing

### 3.1 Construction of Return Series

We choose the Shanghai SE Composite Index (SH000001) and the Shenzhen SE Component Index (SZ399001) these two major indexes in the Chinese stock market as the empirical objects. We collect the daily closing prices from January 5, 2009 to December 31, 2015, which includes 1700 trading days and no public holidays, to form the historical data set of daily returns. Since we are interested in HF returns with 20-minute intervals, 30-minute intervals and 40-minute intervals, we also collect the corresponding data to construct three data sets of intraday returns. Their sizes are 1700×12, 1700×8 and 1700×6, respectively.

The returns in this paper actually refer to logarithmic returns. The daily returns are produced by $R_t = logP_t - logP_t$, and the intraday returns can be obtained by $R_{\tau,t} = logP_{\tau,t} - logP_{\tau-1,t}$, where $P$ refers to the closing price at that time. To maintain the continuity of the variable, the returns at the first moment on day $t$ are defined as the log price at the opening time on day $t$ minus the log price at the closing time on day $t-1$.

From the aspect of risk management, we are most concerned with negative returns caused by extreme events; therefore, we negate all returns and concentrate on the VaR along the right tail. In the following, all so-called returns refer to returns that have been negated; we will not state this point again. Table B1 in Appendix B shows the descriptive statistical results of the historical daily returns of SH000001 and SZ399001 from January 5, 2009 to December 31, 2015, and the unit root test shows that they are all first-order stationary, thereby ensuring the stability of the return series. As mentioned in 2.2, the three GDs are applicable to random variables with nonnegative observations. Therefore, at the beginning of the modeling process, the whole series of observations is shifted to the right by the absolute value of the smallest negative return. The VaR calculated based on the shifted distribution is then be adjusted by subtracting this value.

We then perform Pearson's correlation test to check how the HF data used in the empirical study is correlated with each other. The results indicate a certain degree of correlation between intraday returns. And the dependence of 20 min-HF is particularly significant, while that of 30 min-HF is weakest. Although it implies that the strong-form efficient market hypothesis can be hardly satisfied, we have already considered the seasonal factor to improve the dynamic process of the scale parameter, which could effectively reduce the correlation between HF series. From the empirical results, this practice is quite effective. What's more, to minimize the negative impact of the existing independence, we tend to use low-correlation HF data in the empirical analysis. See the details of test results in Table B2 in Appendix B.

## 3.2  Vacation effect and overnight effect

Generally, the "vacation and weekend effect" refers to the phenomenon that the opening price after the end of a long closed period may be remarkably different from the closing price before the start of the closed period. This abnormal fluctuation may lead to "fake" extreme returns that do not represent actual information about the market. Therefore, returns affected by this effect must be adjusted before constructing models. Specifically, we first remove the special returns from the original data set and then calculate the sample mean and standard deviation. Second, we standardize the returns with the weekend effect and vacation effect separately; finally, we apply inverse standardization for the affected returns by means of the sample mean and standard deviation mentioned above. The HF returns with these effects are processed in the same way.

The "overnight effect" refers to the phenomenon that due to the accumulation of information over night, a significantly different change in stock returns will occur after the stock market opens the next day. Since the overnight effect generally appears in the returns during the first few moments after opening, for these intraday return series, we consider the overnight effect to exist only in the first 20-minute return, the first 30-minute return and the first 40-minute return within each day. Because each initial intraday return contains the overnight effect and they form the corresponding series, this series does not have the heterogeneity caused by this effect. Hence, there is no need to eliminate the overnight effect. In fact, retaining this effect can embody the actual volatility of returns it causes.

### 3.3 Periodic Structure

We assume that the period of seasonal impact on daily returns is 366 days (including the special case of leap years). Since the returns in the empirical study are logarithmic, considering the missing values on holidays and February 29 in nonleap years to be zero is reasonable, which means no price changes occur on those days. Based on historical data, we use the moving average method to decompose the trend and then average the corresponding values on each day in the cycle to obtain the deterministic annual periodic structure. We believe that it is this periodic structure that plays an important role in making the returns at different moment within the day linearly related. So once this deterministic component is added to the driving mechanism of each moment-specific return, the influence of the same structure that may bring about in the volatility of intra-day returns will be neglected.

## 4. Empirical Results

### 4.1 The estimation efficiency of three GD-DCS models

Under the DCS framework, we fit daily returns for the window of January 5, 2009 to December 31, 2015 based on Well-Pareto IV, Gamma-Pareto IV, and Rayleigh-Pareto IV. The log-likelihood, AIC value and running time are used to compare the fitting efficiency of these three distributions. The results in Table 1 show that for both the SH000001 or SZ399001 data sets, for a given sample size, WPD-DCS has the largest ML and the smallest AIC. The running time is also relatively short. The complicated calculation of the gamma function and digamma function may result in GPD-DCS taking considerable time to run. Although RPD-DCS is better than the other two GDs in terms of fitting speed, its log-likelihood is much smaller than that of WPD-DCS and GPD-DCS; moreover, its AIC is much larger than that of the other two. Hence, taking goodness of fit and efficiency into account, WPD-DCS is selected to produce VaR forecasts in the following empirical research.

Table 1. Estimation efficiency of three GDs

| | SH000001 | | | SZ399001 | | |
|---|---|---|---|---|---|---|
| Model | WPD-DCS | GPD-DCS | RPD-DCS | WPD-DCS | GPD-DCS | RPD-DCS |
| The Number of Parameters | 7 | 7 | 7 | 7 | 7 | 7 |
| Running Time (second) | 0.6 | 6.9 | 0.5 | 1.2 | 6.1 | 0.5 |
| Maximum Loglikelihood Value | 2724 | 2104 | 296 | 3891 | 1455 | 181 |
| AIC | -5434 | -4194 | -578 | -7768 | -2896 | -348 |

Note: AIC is the Akaike information criterion (Akaike,1974) calculated as $AIC = 2k - 2\ln(L)$, where $k$ indicates the number of parameters that are the unknown coefficients in the score-driven equation and $L$ indicates the value of the ML function.

### 4.2 In-sample VaR forecasts based on WPD-DCS

The data in the window from January 5, 2009 to December 31, 2014 are used to build the model and obtain the in-sample VaR forecasts. The in-sample results represent the validity of daily returns integrated by intraday returns with different frequencies. This approach helps to identify the relatively better choice of HF data in our application. The coefficients in the driving equations can be estimated through Maximum Likelihood. We only show the estimators of the in-sample 30min HF returns for SH000001 in Table B3 in Appendix B as an example.

For the in-sample VaR generated by the models based on different frequencies of HF, we apply several classic backtesting methods to compare their performance at risk level $\alpha \in \{0.90, 0.91, \ldots, 0.98, 0.99\}$. The results for SH000001 and SZ399001 are listed in Table 2 and Table 3, respectively (to keep the list concise, we present only the four cases where $\alpha$ is 0.92, 0.94, 0.96 and 0.98 in the table; see Appendix B for the complete list). The rank indicates the superiority of these four models under a default level ($\alpha = 0.15$) set in R function, and the p-value indicates whether the null hypothesis of no difference is rejected. The p-value of the relevant test is indicated by the value in parentheses. For SH000001, according to the statistics and the p-values at different levels of $\alpha$, WPD-DCS-VaR based on daily returns is weaker than the 30 min-HF- or 20 min-HF-based models in terms of coverage ability, independence and MCS procedure. This result can be seen as strong evidence that incorporating the information of intraday returns into the model indeed improves the accuracy of VaR forecasts. However, if 40 min HF data are used, the test results are not ideal due to the possible unreasonable sampling frequency, leading to failure to accurately estimate VaR. Furthermore, when $\alpha \leq 0.95$, 20 min HF-based VaR gains an advantage over other HF-based VaR in the LRuc test, LRcc test, DQ test and MCS test, followed by 30 min HF-based VaR. When $\alpha > 0.95$, 30 min HF-based VaR performs best.

Table 2. Backtesting results of in-sample daily VaR estimates, SH000001.

| Model | Alpha | LRuc statistics | LRcc statistics | DQ statistics | MCS rank | Alpha | LRuc statistics | LRcc statistics | DQ statistics | MCS rank |
|---|---|---|---|---|---|---|---|---|---|---|
| VaR-day | 0.92 | **13.83 (0\*\*)** | **13.92 (0\*\*)** | **18.38 (0.01\*)** | 2 | 0.96 | 0.51 (0.48) | 0.51 (0.78) | 9.57 (0.21) | 2 |
| VaR-40minhq | | **140.74 (0\*\*\*)** | **141.14 (0\*\*\*)** | **97.77 (0\*\*\*)** | 4 | | **62.89 (0\*\*\*)** | **63.02 (0\*\*\*)** | **45.77 (0\*\*\*)** | 4 |
| VaR-30minhq | | **12.28 (0\*\*)** | **12.31 (0\*\*)** | **17.84 (0.01\*)** | 3 | | 0.19 (0.66) | 1.73 (0.42) | 13.5 (0.06) | 1 |
| VaR-20minhq | | 0.28 (0.6) | 0.63 (0.73) | 5.08 (0.65) | 1 | | **6.33 (0.01\*)** | **8.13 (0.02\*)** | 13.7 (0.06) | 3 |
| | | | | | 0.16 | | | | | 0.58 |
| VaR-day | 0.94 | **6.65 (0.01\*)** | **7.06 (0.03\*)** | **16.61 (0.02\*)** | 3 | 0.98 | **5.9 (0.02\*)** | 5.96 (0.05) | **20.52 (0\*\*)** | 2 |
| VaR-40minhq | | **99.32 (0\*\*\*)** | **99.59 (0\*\*\*)** | **68.78 (0\*\*\*)** | 4 | | **31.03 (0\*\*\*)** | **31.06 (0\*\*\*)** | **22.66 (0\*\*\*)** | 4 |
| VaR-30minhq | | **5.47 (0.02\*)** | 5.74 (0.06) | 13.36 (0.06) | 2 | | 3.1 (0.08) | 5.72 (0.06) | 13.12 (0.06) | 1 |
| VaR-20minhq | | 1.59 (0.21) | 2.39 (0.3) | 5.66 (0.58) | 1 | | **18.66 (0\*\*)** | **19.4 (0\*\*)** | **26.88 (0\*\*)** | 3 |
| | | | | | 0.62 | | | | | 0.21 |

Note: ∗, ∗∗, and ∗∗∗ represent statistical significance levels of 5%, 1%, and .1%, respectively and bold text indicates rejections at the ∗ probability level.

For SZ399001, 30 min HF-based VaR shows obvious strength in the test results. Except for $\alpha = 0.9$ and $\alpha = 0.99$, the MCS ranking also proves its superiority. The combined test results of the two datasets show the 30 min HF-based-WPD-DCS model presents higher stability and greater performance of VaR forecasts at high risk levels. Table B2 in Appendix B also proves that the intraday return at 30 min intervals has the weakest correlation. Therefore, we believe that 30 min intraday returns are the best modeling object. Therefore, we construct the WPD-DCS model based on 30 min intraday returns to obtain out-of-sample forecasts and compare them with those from other benchmark models.

Table 3. Backtesting results of in-sample daily VaR estimates, SZ399001.

| Model | Alpha | LRuc statistics | LRcc statistics | DQ statistics | MCS rank | Alpha | LRuc statistics | LRcc statistics | DQ statistics | MCS rank |
|---|---|---|---|---|---|---|---|---|---|---|
| VaR-day | 0.92 | **8.83 (0\*\*)** | **8.96 (0.01\*)** | 14.08 (0.05) | 2 | 0.96 | 0.33 (0.57) | 0.8 (0.67) | 8.65 (0.28) | 3 |
| VaR-40minhq | | **13.04 (0\*\*)** | **15.57 (0\*\*)** | **20.52 (0\*\*)** | 3 | | 0.51 (0.48) | 4.27 (0.12) | 11.15 (0.13) | 2 |
| VaR-30minhq | | **6.51 (0.01\*)** | **6.53 (0.04\*)** | 0.09 (1) | 1 | | 0.01 (0.92) | 0.17 (0.92) | 0.04 (1) | 1 |
| VaR-20minhq | | **13.04 (0\*\*)** | **13.55 (0\*\*)** | **22.47 (0\*\*)** | 4 | | 0.72 (0.4) | 2.76 (0.25) | 11.6 (0.11) | 4 |
| | | | | | 0.42 | | | | | 0.47 |
| VaR-day | 0.94 | 3.05 (0.08) | 3.62 (0.16) | **14.4 (0.04\*)** | 3 | 0.98 | 0.03 (0.87) | 0.23 (0.89) | **21.06 (0\*\*)** | 3 |
| VaR-40minhq | | **4.41 (0.04\*)** | **6.46 (0.04\*)** | 11.44 (0.12) | 2 | | 3.1 (0.08) | 5.72 (0.06) | **15.21 (0.03\*)** | 2 |
| VaR-30minhq | | 0.5 (0.48) | 0.56 (0.75) | 0.06 (1) | 1 | | 3.1 (0.08) | 3.1 (0.21) | 0.02 (1) | 1 |
| VaR-20minhq | | 3.05 (0.08) | 3.62 (0.16) | **14.73 (0.04\*)** | 4 | | 3.1 (0.08) | 5.72 (0.06) | **14.82 (0.04\*)** | 4 |
| | | | | | 0.46 | | | | | 0.4 |

Note: *, **, and *** represent statistical significance levels of 5%, 1%, and .1%, respectively and bold text indicates rejections at the * probability level.

Figure 1 intuitively compares the actual returns of SH000001 and the frequency-specific-data-based VaR forecasts. The in-sample VaR based on daily returns can hardly cover the extreme returns, which indicates that WPD-DCS-VaR based on LF may fail to be a qualified risk management tool. In sharp contrast, the VaR of 30 min HF enhances the coverage ability and achieves an overall smaller difference from the actual returns. The fact that the number of real returns covered by the VaR is more in line with expectations makes the 30 min HF-based model perform better in the LRCC and LRUC tests. The chart for SZ399001 is similar to that of SH000001; see Appendix B for the details.

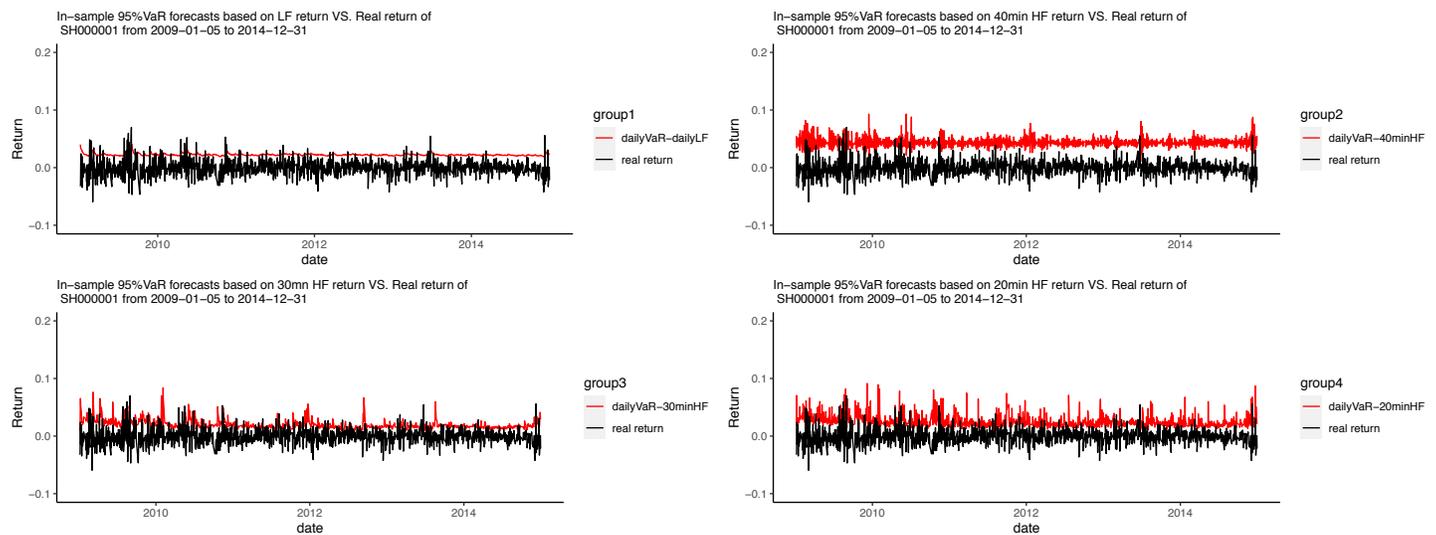

Figure 1. Comparison between frequency-specific return-based VaR forecasts, SH000001.

In the DCS model, since time-varying parameters are driven by a linear combination of the score and past information, whether the scores are autocorrelated is an important indicator for assessing the effectiveness of models. Table B9 in Appendix B lists the p-values of the Lagrange multiplier test for the in-sample scores in WPD-DCS. All p-values are greater than 0.05, indicating that no significant autocorrelation exists. Hence, the established models are proved to be valid.

### 4.3 Out-of-sample VaR forecasts based on WPD-DCS

From May to August 2015, China's stock market experienced two rounds of cliff-shaped declines due to intense negative returns. This abnormal volatility had a substantial impact on the financial industry in a very short period. Therefore, this event is a so-called "stock disaster", and measuring VaR during this period has great empirical significance. We utilize the WPD-DCS model based on intraday data with a 30 min frequency to apply a rolling-window scheme to obtain a time series of VaR forecasts at different confidence levels, i.e., $\alpha \in \{0.90, 0.91, \ldots, 0.98, 0.99\}$. Let $n$ denote the length of VaR to be predicted, $m$ denote the size of the available sample and $s$ denote the length of the rolling window. Then, we have the sequence of forecasts $\{VaR_t^\alpha, t = s+1, s+2 \ldots, s+n\}$, where each prediction is obtained considering the observations that incorporate the intraday return, $\{R_{\tau,t-s}\}_{\tau=1}^8, \{R_{\tau,t-s+1}\}_{\tau=1}^8 \ldots, \{R_{\tau,t-1}\}_{\tau=1}^8$. We produce $n = 244$ daily VaR forecasts from January 5 to December 31 in 2015 by considering the size of the window to be $s = 1456$, and the initial window starts January 5, 2009 and ends December 31, 2014.

Hansen et al. (2012) introduced the RGARCH framework, which combines a GARCH structure for returns with an integrated model for realized measures of volatility. RGARCH offers a substantial improvement in the empirical fit compared to standard GARCH models based on daily returns only. Since our model also incorporates HF financial data to gain a more accurate modeling of daily returns, we consider the RGARCH framework as a benchmark that extends to several kinds of distributions, such as skew-student distribution (SSTD), generalized error distribution (GED) and NIG, and includes realized measures such as realized volatility (RV) and realized range-based volatility (RRV). Since RGARCH models have shown good applicability in many empirical studies, we regard them as benchmarks to assess the performance of out-of-sample daily VaR forecasts generated by 30 min HF-WPD-DCS models.

Table 4 lists the backtesting results along with the MCS ranking of SH000001 at $\alpha \in \{0.93, 0.95, 0.97, 0.99\}$. When $\alpha \geq 0.95$, WPD-DCS-VaR based on 30 min-HF outweighs other RGARCH models in terms of coverage ability. When $\alpha$ is from 0.95 to 0.98, WPD-DCS-VaR always ranks first in the MCS. At these high risk levels, the RGARCH-NIG model with RV as its volatility measurement performs better than other RGARCH models, but it ranks behind WPD-DCS in MCS superiority, which indicates that it is inferior in comprehensive assessment compared with our novel model.

Table 4. Backtesting results of out-of-sample daily VaR forecasts, SH000001.

| | Alpha | LRuc statistics | LRcc statistics | DQ statistics | MCS | Alpha | LRuc statistics | LRcc statistics | DQ statistics | MCS |
|---|---|---|---|---|---|---|---|---|---|---|
| WPD-DCS | | 3.67 (0.06) | 4.35 (0.11) | 13.55 (0.06) | 5 | | 0.37 (0.54) | 1.06 (0.59) | 3.21 (0.86) | 1 |
| RGARCH-SSTD-RV | 0.93 | 0.91 (0.34) | 1.71 (0.43) | 7.24 (0.4) | 1 | 0.97 | 2.6 (0.11) | 2.86 (0.24) | **17.27 (0.02*)** | 3 |
| RGARCH-GED-RV | | 0.51 (0.47) | 1.62 (0.45) | 7.9 (0.34) | 3 | | 1.66 (0.2) | 2.1 (0.35) | **17.28 (0.02*)** | 5 |
| RGARCH-NIG-RV | | 0.51 (0.47) | 1.62 (0.45) | 7.93 (0.34) | 4 | | 1.66 (0.2) | 2.1 (0.35) | **17.32 (0.02*)** | 2 |

| Model | | | | | | | | | | |
|---|---|---|---|---|---|---|---|---|---|---|
| RGARCH-SSTD-RRV | | 1.41 (0.24) | 5.59 (0.06) | **24.4 (0\*\*)** | 7 | | **4.99 (0.03\*)** | **6.46 (0.04\*)** | **16.48 (0.02\*)** | 6 |
| RGARCH-GED-RRV | | 2 (0.16) | 5.52 (0.06) | **22.02 (0\*\*)** | 2 | | 3.71 (0.05) | 3.84 (0.15) | **19.69 (0.01\*)** | 4 |
| RGARCH-NIG-RRV | | 1.41 (0.24) | 5.59 (0.06) | **24.29 (0\*\*)** | 6 | | 3.71 (0.05) | 3.84 (0.15) | **18.25 (0.01\*)** | 7 |
| | | | | | 0.26 | | | | | 0.32 |
| WPD-DCS | | 0.44 (0.51) | 1.12 (0.57) | 10.53 (0.16) | 1 | | **8.01 (0\*\*)** | **8.55 (0.01\*)** | 2.06 (0.96) | 2 |
| RGARCH-SSTD-RV | | 1.14 (0.29) | 4.07 (0.13) | 9.88 (0.2) | 3 | | **8.01 (0\*\*)** | **8.55 (0.01\*)** | **54.91 (0\*\*\*)** | 4 |
| RGARCH-GED-RV | | 1.14 (0.29) | 4.07 (0.13) | 9.98 (0.19) | 4 | | **8.01 (0\*\*)** | **8.55 (0.01\*)** | **55.09 (0\*\*)** | 5 |
| RGARCH-NIG-RV | 0.95 | 0.63 (0.43) | 1.74 (0.42) | 5.65 (0.58) | 2 | 0.99 | 5.72 (0.02) | 6.14 (0.05) | **20.47 (0\*\*)** | 1 |
| RGARCH-SSTD-RRV | | 3.44 (0.06) | 4.9 (0.09) | **16.85 (0.02\*)** | 6 | | **10.55 (0\*\*)** | **11.25 (0\*\*)** | **55.7 (0\*\*\*)** | 7 |
| RGARCH-GED-RRV | | 2.55 (0.11) | 4.44 (0.11) | 10.08 (0.18) | 5 | | **8.01 (0\*\*)** | **8.55 (0.01\*)** | **54.74 (0\*\*\*)** | 3 |
| RGARCH-NIG-RRV | | 3.44 (0.06) | 4.9 (0.09) | **16.65 (0.02\*)** | 7 | | **8.01 (0\*\*)** | **8.55 (0.01\*)** | **55.79 (0\*\*\*)** | 6 |
| | | | | | 0.45 | | | | | 0.63 |

Note: ∗, ∗∗, and ∗∗∗ represent statistical significance levels of 5%, 1%, and .1%, respectively and bold text indicates rejections at the * probability level.

Table 5 gives the evaluation results of SZ399001. In contrast with SH000001, the WPD-DCS-VaR based on SZ3990001 shows an obvious advantage over RGARCH models in terms of coverage and MCS. Specifically, when $\alpha \geq 0.95$, WPD-DCS-VaR ranks first in the MCS assessment, and when $\alpha < 0.95$, WPD-DCS-VaR maintains a position in the top three. Similarly, on this empirical dataset, RGARCH-NIG-RV is also the best RGARCH model.

Table 5. Backtesting results of out-of-sample daily VaR forecasts, SZ399001.

| Model | Alpha | LRuc statistics | LRcc statistics | DQ statistics | MCS | Alpha | LRuc statistics | LRcc statistics | DQ statistics | MCS |
|---|---|---|---|---|---|---|---|---|---|---|
| WPD-DCS | 0.93 | 0.28 (0.59) | 3.85 (0.15) | **22.34 (0\*\*)** | 2 | 0.97 | 3.71 (0.05) | 5.62 (0.06) | **45.23 (0\*\*)** | 1 |
| RGARCH-SSTD-RV | | 2 (0.16) | 5.52 (0.06) | 13.45 (0.06) | 6 | | **4.99 (0.03\*)** | **6.46 (0.04\*)** | 14.04 (0.05) | 5 |
| RGARCH-GED-RV | | 1.41 (0.24) | 5.59 (0.06) | 13.15 (0.07) | 1 | | **4.99 (0.03\*)** | **6.46 (0.04\*)** | 14 (0.05) | 3 |
| RGARCH-NIG-RV | | 1.41 (0.24) | 5.59 (0.06) | 10.84 (0.15) | 4 | | **4.99 (0.03\*)** | **6.46 (0.04\*)** | 14.05 (0.05) | 2 |
| RGARCH-SSTD-RRV | | **5.33 (0.02\*)** | **13.4 (0\*\*)** | **33.42 (0\*\*)** | 5 | | **7.99 (0\*\*)** | **14.05 (0\*\*)** | **34.29 (0\*\*)** | 7 |
| RGARCH-GED-RRV | | **5.33 (0.02\*)** | **13.4 (0\*\*)** | **33.44 (0\*\*)** | 7 | | **9.69 (0\*\*)** | **18.5 (0\*\*)** | **50.47 (0\*\*\*)** | 6 |
| RGARCH-NIG-RRV | | **5.33 (0.02\*)** | **13.4 (0\*\*)** | **33.45 (0\*\*)** | 3 | | **6.42 (0.01\*)** | **9.98 (0.01\*)** | **25.24 (0\*\*)** | 4 |
| | | | | | 0.52 | | | | | 0.17 |
| WPD-DCS | 0.95 | 0.27 (0.61) | 4.56 (0.1) | **31.03 (0\*\*)** | 1 | 0.99 | **4.9 (0.03\*)** | 4.9 (0.09) | 2.42 (0.93) | 1 |
| RGARCH-SSTD-RV | | **4.44 (0.04\*)** | **7.49 (0.02\*)** | 13.56 (0.06) | 5 | | **10.55 (0\*\*)** | **11.25 (0\*\*)** | **29.36 (0\*\*)** | 5 |
| RGARCH-GED-RV | | 3.44 (0.06) | **7.12 (0.03\*)** | 13.05 (0.07) | 2 | | **10.55 (0\*\*)** | **11.25 (0\*\*)** | **28.9 (0\*\*)** | 3 |
| RGARCH-NIG-RV | | 3.44 (0.06) | **7.12 (0.03\*)** | 13.04 (0.07) | 3 | | **8.01 (0\*\*)** | **8.55 (0.01\*)** | **27.31 (0\*\*)** | 2 |
| RGARCH-SSTD-RRV | | **9.49 (0\*\*)** | **21.08 (0\*\*)** | **54.58 (0\*\*\*)** | 7 | | **13.33 (0\*\*)** | **14.19 (0\*\*)** | **63.01 (0\*\*)** | 7 |
| RGARCH-GED-RRV | | **9.49 (0\*\*)** | **21.08 (0\*\*)** | **54.77 (0\*\*\*)** | 6 | | **19.49 (0\*\*)** | **19.76 (0\*\*)** | **74.62 (0\*\*)** | 6 |
| RGARCH-NIG-RRV | | **6.76 (0.01\*)** | **13.78 (0\*\*)** | **26.86 (0\*\*)** | 4 | | **10.55 (0\*\*)** | **11.25 (0\*\*)** | **59.67 (0\*\*)** | 4 |
| | | | | | 0.38 | | | | | 0.23 |

Note: ∗, ∗∗, and ∗∗∗ represent statistical significance levels of 5%, 1%, and .1%, respectively and bold text

indicates rejections at the * probability level.

For both SH000001 and SZ399001, at a high risk level of α≥0.95, WPD-DCS-VaR has a prominent advantage compared with other RGARCH-VaRs. Therefore, the out-of-sample VaR forecasts of both WPD-DCS and RGARCH under $\alpha \in \{0.96, 0.97, 0.98\}$ are visually compared in Figure 2. When clusters of extreme returns occur, for example, in mid-June and late August, WPD-DCS-VaR can sensitively capture the fluctuations in returns and cover the actual returns in a more accurate manner due to its unique DCS mechanism. The out-of-sample scores also pass the LM test. Therefore, there is no significant serial autocorrelation in the scores generated by WPD-DCS based on the 30-min returns. The details are provided in Appendix B.

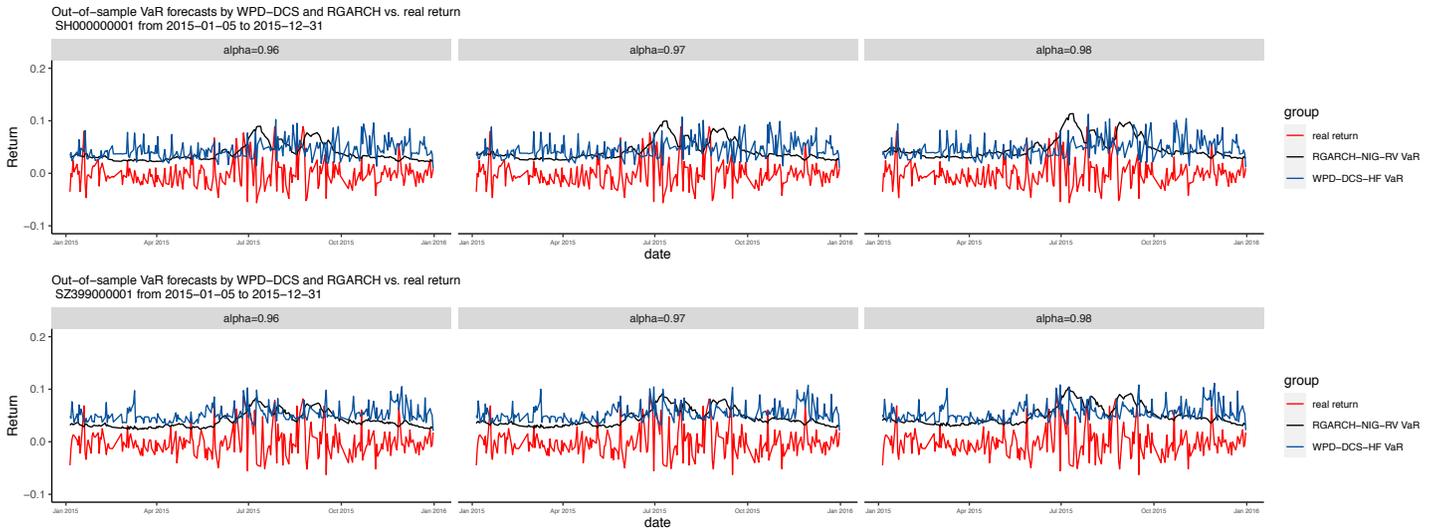

Figure 2. Comparison of out-of-sample VaR forecasts between the 30 min-WPD-DCS-based and RGARCH-based methods.

The fact that the novel model proposed in this study could improve VaR measurement lies in the out-of-sample results. The evidence to support this result is that when $\alpha$ is high, e.g., when $\alpha \geq 0.95$, the WPD-DCS model based on HF data provides more accurate VaR predictions, thereby avoiding the serious consequences caused by underestimating the risk. The appropriate form of the GD specified for returns and the incorporation of HF data, along with the delicate mechanism of estimating the dynamic parameters, are three indispensable parts of this approach. Hence, we believe this novel tool promotes the accuracy of VaR forecasts and thus contributes to financial risk management.

## 5. Conclusion

The purpose of this paper is to develop a new parametric approach to improving VaR forecasts and thus contribute to risk management. We propose the GD-DCS-VaR model, which is a method of directly using intraday returns to estimate the conditional distribution of daily returns. Daily returns can be obtained by summing the intraday returns estimated by a generalized-distribution-based model driven by the conditional score. The bootstrap method makes the simulation of daily returns feasible, and the empirical analysis demonstrates the effectiveness and tractability of this novel approach.

Overall, our study provides a complete and standard paradigm for improving the parametric VaR prediction model, enriching the theoretical understanding of risk management tools. First, we reproduce the derivation process of a class of GDs. Second, we construct a DCS framework on these GDs to emphasize its practicability. Third, we highlight the contribution of intraday information in fitting daily returns and predicting daily VaR. Thus, the accurate estimation of intraday return is likely to be key to improving risk management tools. In the future, we expect to extend this line of thinking and to solve tough problems, such as dealing with intraday returns that are seriously correlated. We may consider the GD-DCS-VaR model that incorporates copula to give a more applicable scheme.

## Declaration of interests

The authors report no conflicts of interest. The authors alone are responsible for the content and writing of the paper.

**Appendix A**
1. $X \sim WPD^{IV}(\alpha, \beta, c)$

$$\nabla_\beta = \frac{\partial \ln g(x)}{\partial \beta} = \frac{c}{\beta} - c\left[\beta \ln\left(1 + x^{\frac{1}{\alpha}}\right)\right]^{c-1} \ln\left(1 + x^{\frac{1}{\alpha}}\right)$$

$$\nabla_c = \frac{\partial lng(x)}{\partial c} = \frac{1}{c} + ln\beta + \ln\left[\ln\left(1+x^{\frac{1}{\alpha}}\right)\right] - \ln\left[\beta\ln\left(1+x^{\frac{1}{\alpha}}\right)\right] \cdot \left[\beta\ln\left(1+x^{\frac{1}{\alpha}}\right)\right]^c$$

$$\frac{\partial^2 lng(x)}{\partial \beta^2} = -\frac{c}{\beta^2} - c(c-1)\left[\beta\ln\left(1+x^{\frac{1}{\alpha}}\right)\right]^{c-2} \cdot \left[\ln\left(1+x^{\frac{1}{\alpha}}\right)\right]^2$$

$$\frac{\partial^2 lng(x)}{\partial c^2} = -\frac{1}{c^2} - \{\ln[\beta\ln(1+x^{\frac{1}{\alpha}})]\}^2 \cdot \left[\beta\ln\left(1+x^{\frac{1}{\alpha}}\right)\right]^c$$

We can now compute the conditional expectation. First note that

$$E\left(\frac{\partial lng(x)}{\partial \beta}\right) = 0 \Rightarrow E\left(\left[\ln\left(1+x^{\frac{1}{\alpha}}\right)\right]^c\right) = \beta^{-c},$$

implying that

$$E\left(\frac{\partial^2 lng(x)}{\partial \beta^2}\right) = -\frac{c}{\beta^2} - c(c-1)\beta^{c-2}\beta^{-c} = -\left(\frac{c}{\beta}\right)^2$$

$$E\left(\frac{\partial^2 lng(x)}{\partial c^2}\right) = -\frac{1}{c^2} - \int_0^{+\infty} c\ln^2(m) \cdot m^{2c-1}e^{-m^c}dm = -\frac{1}{c^2} - \lim_{n\to\infty}\int_0^n c\ln^2(m) \cdot m^{2c-1}e^{-m^c}dm$$

$$= -\frac{1}{c^2} - \lim_{n\to\infty} c\ln^2(m) \cdot m^{2c-1}e^{-m^c}$$

Thus, if $\frac{1}{E\left(\frac{\partial^2 lng(x)}{\partial c^2}\right)}$ acts as the factor to standardize the score, equation (18) in the body will degenerate to $v_t = A_2 + B_2 v_{t-1}$, i.e., the score of $v$ losses its role of driving.

2. $X \sim GPD^{IV}(\theta, \alpha, c)$

$$\nabla_c = \frac{\partial lng(x)}{\partial c} = -\frac{\theta}{c} + \frac{1}{c^2}\ln\left(1+x^{\frac{1}{\alpha}}\right)$$

$$\nabla_\theta = \frac{\partial lng(x)}{\partial \theta} = -\ln(c) - \psi(\theta) + \ln\left[\ln\left(1+x^{\frac{1}{\alpha}}\right)\right],$$

where $\psi(\theta) = \frac{\Gamma'(\theta)}{\Gamma(\theta)}$ that denotes the digamma function.

Note that

$$E\left(\frac{\partial lng(x)}{\partial c}\right) = 0 \Rightarrow E\left(\ln\left(1+x^{\frac{1}{\alpha}}\right)\right) = c\theta,$$

then

$$E\left(\frac{\partial^2 lng(x)}{\partial c^2}\right) = \frac{-\theta}{c^2},$$

implying that

$$E\left(\frac{\partial^2 lng(x)}{\partial\theta^2}\right) = -\psi'(\theta)$$

3. $X \sim RPD^{IV}(\sigma, \alpha, \delta)$

$$\nabla_\sigma = \frac{\partial lng(x)}{\partial \sigma} = -\frac{2}{\sigma} + \frac{\delta^2}{\sigma^3}\left[\ln\left(1 + x^{\frac{1}{\alpha}}\right)\right]^2$$

$$\nabla_\delta = \frac{\partial lng(x)}{\partial \delta} = \frac{2}{\delta} - \frac{\delta\left[\ln\left(1 + x^{\frac{1}{\alpha}}\right)\right]^2}{\sigma^2}$$

Note that

$$E\left(\frac{\partial lng(x)}{\partial \sigma}\right) = 0 \text{ or } E\left(\frac{\partial lng(x)}{\partial \delta}\right) = 0 \Rightarrow E\left(\left[\ln\left(1 + x^{\frac{1}{\alpha}}\right)\right]^2\right) = \frac{2\sigma^2}{\delta^2},$$

implying that

$$E\left(\frac{\partial^2 lng(x)}{\partial \sigma^2}\right) = \frac{2}{\sigma^2} - \frac{3\delta^2}{\sigma^4} E\left(\left[\ln\left(1 + x^{\frac{1}{\alpha}}\right)\right]^2\right) = \frac{-4}{\sigma^2}$$

$$E\left(\frac{\partial^2 lng(x)}{\partial \delta^2}\right) = \frac{-2}{\delta^2} - \frac{1}{\sigma^2} E\left(\left[\ln\left(1 + x^{\frac{1}{\alpha}}\right)\right]^2\right) = \frac{-4}{\delta^2}$$

**Appendix B**

Table B1. Descriptive statistics of the daily returns of H000001 and SZ399001

| | sh000001 $p_t$ | sh000001 $ln(p_t/p_{t-1})$ | sz399001 $p_t$ | sz399001 $ln(p_t/p_{t-1})$ |
|---|---|---|---|---|
| Start date | 2009-01-05 | 2009-01-05 | 2009-01-05 | 2009-01-05 |
| End date | 2015-12-31 | 2015-12-31 | 2015-12-31 | 2015-12-31 |
| Sample size | 1700 | 1700 | 1700 | 1700 |
| Minimum | 1880.72 | -0.0815 | 6634.88 | -0.1103 |
| Maximum | 5178.19 | 0.0547 | 18211.76 | 0.05701 |
| Average | 2685.763 | -1e-04 | 10502.45 | -1e-04 |
| Standard deviation | 603.0049 | 0.0128 | 2160.589 | 0.0151 |
| Skewness | 1.3288 | -0.7813 | 0.4623 | -0.8916 |
| kurtosis | 2.0434 | 4.6337 | -0.1824 | 5.4299 |
| ADF statistic | -1.783(0.6703) | -8.2203(<0.01**) | -2.4815(0.3745) | -11.595(<0.01**) |

Note: *, **, and *** represent statistical significance at the 5%, 1%, and .1% levels, respectively. The p-value of the relevant test is indicated by the value in parentheses.

Table B2. Results of Pearson's correlation test for intraday returns, SH000001 and SZ399001, 2009–2015.

| Pair of returns | SH000001 | | | SZ399001 | | |
|---|---|---|---|---|---|---|
| | 20min-HF | 30min-HF | 40min-HF | 20min-HF | 30min-HF | 40min-HF |
| 1-2   | 0(***)  | 0.89    | 0.98    | 0(***)  | 0.52    | 0.78    |
| 2-3   | 0(***)  | 0(***)  | 0.34    | 0(***)  | 0.14    | 0.14    |
| 3-4   | 0(***)  | 0.27    | 0(***)  | 0.3     | 0(***)  | 0(***)  |
| 4-5   | 0.57    | 0(***)  | 0(***)  | 0(***)  | 0(***)  | 0(***)  |
| 5-6   | 0(***)  | 0.55    | 0(***)  | 0(***)  | 0.85    | 0(***)  |
| 6-7   | 0(***)  | 0(***)  |         | 0(***)  | 0(***)  |         |
| 7-8   | 0(***)  | 0(***)  |         | 0.28    | 0(***)  |         |
| 8-9   | 0(***)  |         |         | 0.01(**)|         |         |
| 9-10  | 0.36    |         |         | 0.34    |         |         |
| 10-11 | 0(***)  |         |         | 0(***)  |         |         |
| 11-12 | 0(***)  |         |         | 0(***)  |         |         |
| Ratio | 81.82%  | 57.14%  | 60.00%  | 72.73%  | 57.14%  | 60.00%  |

Note: *, **, and *** represent statistical significance at the 5%, 1%, and .1% levels, respectively. The values in the table are the p-values of the Pearson's correlation test, which determines whether to reject the null hypothesis that the correlation between two series is equal to 0. Ratio gives the proportion of sequence pairs that do not meet the independence requirement under each frequency.

Table B3. The estimated coefficients in $\theta$ based on in-sample 30min HF data, SH000001.

| $\tau$ | $A_1$ | $B_1$ | $C_1$ | $A_2$ | $B_2$ | $C_2$ |
|---|---|---|---|---|---|---|
| 1 | -0.11 (0.33) | -0.09 (0.01) | 1.02 (0.22) | 1.66 (0.49) | -0.42 (0.39) | -0.05 (0.38) |
| 2 | 0.23 (0.06) | -0.01 (0.03) | 0.98 (0.36) | 0.92 (0.36) | 0.62 (0.46) | -1.29 (0.54) |
| 3 | 0.49 (0.21) | -2.75 (0.34) | 0.69 (0.04) | 3.27 (0.85) | -0.61 (0.33) | -1.14 (0.09) |
| 4 | -0.61 (0.12) | -1.88 (0.26) | 0.95 (0.18) | -0.03 (0.07) | 0.61 (0.28) | 0.93 (0.03) |
| 5 | 0.72 (0.02) | -1.55 (0.09) | 0.51 (0.41) | -5.37 (0.28) | -0.18 (0.73) | 3.75 (0.08) |
| 6 | 0.45 (0.08) | -2.86 (0.43) | 0.27 (0.06) | 2.99 (0.31) | -0.23 (0.63) | -1.52 (0.53) |
| 7 | 0.25 (0.04) | -3.08 (0.51) | 0.78 (0.62) | 2.99 (0.69) | -0.72 (0.32) | -1.24 (0.09) |
| 8 | 0.14 (0.11) | -0.98 (0.46) | 0.97 (0.33) | 0.12 (0.47) | 0.21 (0.66) | 0.86 (0.13) |

Note: values in parentheses indicate the standard error of the estimated results

Table B4. Back testing results of in-sample daily VaR estimates, SH000001.

| Model | Alpha | LRuc statistics | LRcc | DQ statistics | MCS | Alpha | LRuc | LRcc statistics | DQ statistics | MCS |
|---|---|---|---|---|---|---|---|---|---|---|

|  |  | statistics |  |  | rank |  | statistics |  |  | rank |
|---|---|---|---|---|---|---|---|---|---|---|
| VaR-day | 0.9 | **13.09 (0**)** | **13.1 (0**)** | **18.42 (0.01*)** | 2 | 0.95 | 1.77 (0.18) | 1.82 (0.4) | 13.4 (0.06) | 2 |
| VaR-40minhq |  | **160.21 (0***)** | **161.16 (0***)** | **117.07 (0***)** | 4 |  | **84.74 (0***)** | **84.91 (0***)** | **59.63 (0***)** | 4 |
| VaR-30minhq |  | **18.46 (0**)** | **18.47 (0**)** | **25.15 (0**)** | 3 |  | 2.51 (0.11) | 3.39 (0.18) | 11.32 (0.13) | 1 |
| VaR-20minhq |  | 0.72 (0.4) | 0.87 (0.65) | 4.41 (0.73) | 1 |  | 3.99 (0.05) | 6.06 (0.05) | 10.49 (0.16) | 3 |
|  |  |  |  |  | 0.6 |  |  |  |  | 0.63 |
| VaR-day | 0.91 | **13.38 (0**)** | **14.12 (0**)** | **19 (0.01*)** | 2 | 0.96 | 0.51 (0.48) | 0.51 (0.78) | 9.57 (0.21) | 2 |
| VaR-40minhq |  | **144.62 (0***)** | **145.36 (0***)** | **106.39 (0***)** | 4 |  | **62.89 (0***)** | **63.02 (0***)** | **45.77 (0***)** | 4 |
| VaR-30minhq |  | **15.71 (0**)** | **15.78 (0**)** | **23.25 (0**)** | 3 |  | 0.19 (0.66) | 1.73 (0.42) | 13.5 (0.06) | 1 |
| VaR-20minhq |  | 0.14 (0.71) | 0.23 (0.89) | 4.45 (0.73) | 1 |  | **6.33 (0.01*)** | **8.13 (0.02*)** | 13.7 (0.06) | 3 |
|  |  |  |  |  | 0.6 |  |  |  |  | 0.58 |
| VaR-day | 0.92 | **13.83 (0**)** | **13.92 (0**)** | **18.38 (0.01*)** | 2 | 0.97 | 0.64 (0.42) | 0.96 (0.62) | 13.22 (0.07) | 2 |
| VaR-40minhq |  | **140.74 (0***)** | **141.14 (0***)** | **97.77 (0***)** | 4 |  | **52.53 (0***)** | **52.58 (0***)** | **34.95 (0***)** | 4 |
| VaR-30minhq |  | **12.28 (0**)** | **12.31 (0**)** | **17.84 (0.01*)** | 3 |  | 0.25 (0.61) | 3.32 (0.19) | 13.79 (0.05) | 1 |
| VaR-20minhq |  | 0.28 (0.6) | 0.63 (0.73) | 5.08 (0.65) | 1 |  | **15.9 (0**)** | **15.96 (0**)** | **22.57 (0**)** | 3 |
|  |  |  |  |  | 0.16 |  |  |  |  | 0.4 |
| VaR-day | 0.93 | **11.99 (0**)** | **12.11 (0**)** | **20.2 (0.01*)** | 3 | 0.98 | **5.9 (0.02*)** | 5.96 (0.05) | **20.52 (0**)** | 2 |
| VaR-40minhq |  | **114.16 (0***)** | **114.57 (0***)** | **82.32 (0***)** | 4 |  | **31.03 (0***)** | **31.06 (0***)** | **22.66 (0***)** | 4 |
| VaR-30minhq |  | **8.37 (0**)** | **8.71 (0.01*)** | **16.79 (0.02*)** | 2 |  | 3.1 (0.08) | 5.72 (0.06) | 13.12 (0.06) | 1 |
| VaR-20minhq |  | 0.38 (0.54) | 1.56 (0.46) | 6.59 (0.47) | 1 |  | **18.66 (0**)** | **19.4 (0**)** | **26.88 (0**)** | 3 |
|  |  |  |  |  | 0.39 |  |  |  |  | 0.21 |
| VaR-day | 0.94 | **6.65 (0.01*)** | **7.06 (0.03*)** | **16.61 (0.02*)** | 3 | 0.99 | **9.87 (0**)** | **10.2 (0.01*)** | **37.78 (0**)** | 3 |
| VaR-40minhq |  | **99.32 (0***)** | **99.59 (0***)** | **68.78 (0***)** | 4 |  | **13.73 (0**)** | **13.75 (0**)** | 13.03 (0.07) | 1 |
| VaR-30minhq |  | **5.47 (0.02*)** | 5.74 (0.06) | 13.36 (0.06) | 2 |  | **12.66 (0**)** | **12.87 (0**)** | **20.25 (0**)** | 2 |
| VaR-20minhq |  | 1.59 (0.21) | 2.39 (0.3) | 5.66 (0.58) | 1 |  | **39.05 (0**)** | **41.79 (0**)** | **73.15 (0***)** | 4 |
|  |  |  |  |  | 0.62 |  |  |  |  | 0.12 |

Note: ∗, ∗∗, and ∗∗∗ represent statistical significance levels of 5%, 1%, and .1%, respectively. The rank tells the superiority of these four models under a default level ($\alpha = 0.15$) set in R function, and p-value helps to prove the non-rejection of this superiority. The p-value of the relevant test is indicated by the value in parentheses. Bold text indicates rejections at the * probability level.

Table B5. Back testing results of in-sample daily VaR estimates, SZ399001.

| Model | Alpha | LRuc statistics | LRcc statistics | DQ statistics | MCS rank | Alpha | LRuc statistics | LRcc statistics | DQ statistics | MCS rank |
|---|---|---|---|---|---|---|---|---|---|---|
| VaR-day | 0.9 | **11.1 (0**)** | **11.1 (0**)** | **14.47 (0.04*)** | 1 | 0.95 | 0.69 (0.41) | 1.94 (0.38) | 10.78 (0.15) | 3 |
| VaR-40minhq |  | **17.63 (0**)** | **21.35 (0**)** | **31.75 (0**)** | 3 |  | 1.45 (0.23) | 3.14 (0.21) | 8.79 (0.27) | 2 |
| VaR-30minhq |  | **16.82 (0**)** | **16.82 (0**)** | 0.11 (1) | 2 |  | 0.21 (0.64) | 0.24 (0.89) | 0.05 (1) | 1 |
| VaR-20minhq |  | **20.19 (0**)** | **20.58 (0**)** | **30.16 (0**)** | 4 |  | 2.13 (0.14) | 2.89 (0.24) | 9.03 (0.25) | 4 |
|  |  |  |  |  | 0.34 |  |  |  |  | 0.58 |
| VaR-day | 0.91 | **10.59 (0**)** | **10.64 (0**)** | **15.75 (0.03*)** | 2 | 0.96 | 0.33 (0.57) | 0.8 (0.67) | 8.65 (0.28) | 3 |
| VaR-40minhq |  | **17.39 (0**)** | **19.83 (0**)** | **29.45 (0**)** | 4 |  | 0.51 (0.48) | 4.27 (0.12) | 11.15 (0.13) | 2 |
| VaR-30minhq |  | **10.59 (0**)** | **10.64 (0**)** | 0.1 (1) | 1 |  | 0.01 (0.92) | 0.17 (0.92) | 0.04 (1) | 1 |

| | | | | | | | | | |
|---|---|---|---|---|---|---|---|---|---|
| VaR-20minhq | | 17.39 (0**) | 17.94 (0**) | 28.14 (0**) | 3 | | 0.72 (0.4) | 2.76 (0.25) | 11.6 (0.11) | 4 |
| | | | | | 0.44 | | | | | 0.47 |
| VaR-day | 0.92 | 8.83 (0**) | 8.96 (0.01*) | 14.08 (0.05) | 2 | 0.97 | 0.07 (0.8) | 0.11 (0.95) | 17.13 (0.02*) | 3 |
| VaR-40minhq | | 13.04 (0**) | 15.57 (0**) | 20.52 (0**) | 3 | | 0.04 (0.84) | 3.59 (0.17) | 13.05 (0.07) | 2 |
| VaR-30minhq | | 6.51 (0.01*) | 6.53 (0.04*) | 0.09 (1) | 1 | | 0.25 (0.61) | 0.4 (0.82) | 0.03 (1) | 1 |
| VaR-20minhq | | 13.04 (0**) | 13.55 (0**) | 22.47 (0**) | 4 | | 0.04 (0.84) | 3.59 (0.17) | 10.7 (0.15) | 4 |
| | | | | | 0.42 | | | | | 0.48 |
| VaR-day | 0.93 | 6.53 (0.01*) | 6.7 (0.04*) | 14.01 (0.05) | 2 | 0.98 | 0.03 (0.87) | 0.23 (0.89) | 21.06 (0**) | 3 |
| VaR-40minhq | | 7.12 (0.01*) | 10.5 (0.01*) | 15.11 (0.03*) | 3 | | 3.1 (0.08) | 5.72 (0.06) | 15.21 (0.03*) | 2 |
| VaR-30minhq | | 2.81 (0.09) | 2.82 (0.24) | 0.08 (1) | 1 | | 3.1 (0.08) | 3.1 (0.21) | 0.02 (1) | 1 |
| VaR-20minhq | | 8.37 (0**) | 8.71 (0.01*) | 20.51 (0**) | 4 | | 3.1 (0.08) | 5.72 (0.06) | 14.82 (0.04*) | 4 |
| | | | | | 0.48 | | | | | 0.4 |
| VaR-day | 0.94 | 3.05 (0.08) | 3.62 (0.16) | 14.4 (0.04*) | 3 | 0.99 | 6.23 (0.01*) | 6.8 (0.03*) | 63.32 (0***) | 1 |
| VaR-40minhq | | 4.41 (0.04*) | 6.46 (0.04*) | 11.44 (0.12) | 2 | | 12.66 (0**) | 14.73 (0**) | 30.66 (0**) | 3 |
| VaR-30minhq | | 0.5 (0.48) | 0.56 (0.75) | 0.06 (1) | 1 | | 9.87 (0**) | 10.96 (0**) | 0.01 (1) | 2 |
| VaR-20minhq | | 3.05 (0.08) | 3.62 (0.16) | 14.73 (0.04*) | 4 | | 14.16 (0**) | 16.04 (0**) | 46.08 (0**) | 4 |
| | | | | | 0.46 | | | | | 0.43 |

Note: ∗, ∗∗, and ∗∗∗ represent statistical significance levels of 5%, 1%, and .1%, respectively. The rank tells the superiority of these four models under a default level ($\alpha = 0.15$) set in R function, and p-value helps to prove the non-rejection of this superiority. The p-value of the relevant test is indicated by the value in parentheses. Bold text indicates rejections at the * probability level.

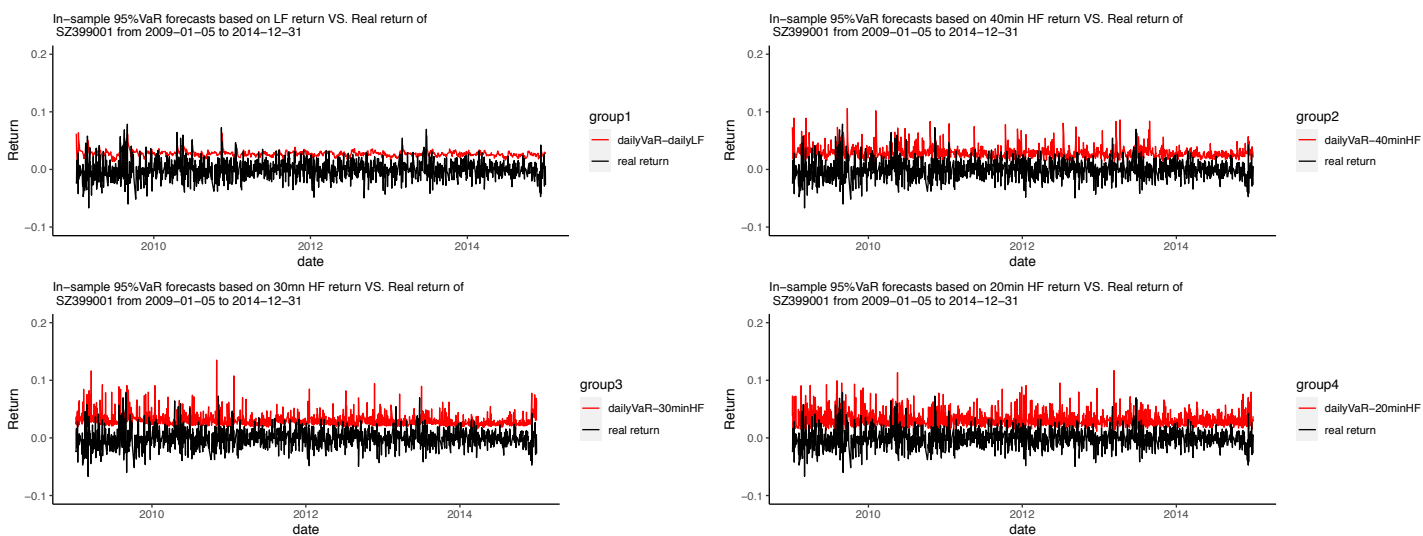

Figure B1. Comparison between frequency-specific-return-based VaR forecasts, SZ399001.

Table B6. P-values of LM test for in-sample parameter's scores of 30min return.

| | SH000001-lambda | SZ399001-lambda |
|---|---|---|
| 1 | 0.17 | 0.96 |
| 2 | 0.23 | 0.13 |
| 3 | 0.71 | 0.22 |
| 4 | 0.98 | 0.47 |
| 5 | 0.89 | 0.31 |
| 6 | 0.13 | 0.35 |
| 7 | 0.37 | 0.3 |
| 8 | 0.28 | 0.92 |

Note: a p-value greater than 0.05 indicates that the null hypothesis that the sequence is not self-correlated is not rejected.

Table B7. Back testing results of out-of-sample daily VaR forecasts, SH000001.

| Model | Alpha | LRuc statistics | LRcc statistics | DQ statistics | MCS | Alpha | LRuc statistics | LRcc statistics | DQ statistics | MCS |
|---|---|---|---|---|---|---|---|---|---|---|
| WPD-DCS |  | **11.88 (0\*\*)** | **12.56 (0\*\*)** | 0.11 (1) | 7 |  | 0.44 (0.51) | 1.12 (0.57) | 10.53 (0.16) | 1 |
| RGARCH-SSTD-RV |  | 0.3 (0.58) | 3.45 (0.18) | **16.47 (0.02\*)** | 2 |  | 1.14 (0.29) | 4.07 (0.13) | 9.88 (0.2) | 3 |
| RGARCH-GED-RV |  | 0.3 (0.58) | 3.45 (0.18) | **16.48 (0.02\*)** | 3 |  | 1.14 (0.29) | 4.07 (0.13) | 9.98 (0.19) | 4 |
| RGARCH-NIG-RV | 0.9 | 0.3 (0.58) | 3.45 (0.18) | **16.49 (0.02\*)** | 4 | 0.95 | 0.63 (0.43) | 1.74 (0.42) | 5.65 (0.58) | 2 |
| RGARCH-SSTD-RRV |  | 1.34 (0.25) | **9.17 (0.01\*)** | **21.4 (0\*\*)** | 6 |  | 3.44 (0.06) | 4.9 (0.09) | **16.85 (0.02\*)** | 6 |
| RGARCH-GED-RRV |  | 0.91 (0.34) | 3 (0.22) | **18.33 (0.01\*)** | 1 |  | 2.55 (0.11) | 4.44 (0.11) | 10.08 (0.18) | 5 |
| RGARCH-NIG-RRV |  | 1.34 (0.25) | **9.17 (0.01\*)** | **21.32 (0\*\*)** | 5 |  | 3.44 (0.06) | 4.9 (0.09) | **16.65 (0.02\*)** | 7 |
|  |  |  |  |  | 0.18 |  |  |  |  | 0.45 |
| WPD-DCS |  | **8.82 (0\*\*)** | **9.49 (0.01\*)** | **17.71 (0.01\*)** | 7 |  | 0.06 (0.8) | 0.76 (0.69) | 9.48 (0.22) | 1 |
| RGARCH-SSTD-RV |  | 0.44 (0.5) | 2.81 (0.25) | **15.5 (0.03\*)** | 2 |  | 2.53 (0.11) | 3.64 (0.16) | 8.95 (0.26) | 3 |
| RGARCH-GED-RV |  | 0.44 (0.5) | 2.81 (0.25) | **15.49 (0.03\*)** | 3 |  | 2.53 (0.11) | 3.64 (0.16) | 9.12 (0.24) | 5 |
| RGARCH-NIG-RV | 0.91 | 0.44 (0.5) | 2.81 (0.25) | **15.49 (0.03\*)** | 4 | 0.96 | 1.7 (0.19) | 3.17 (0.2) | 8.88 (0.26) | 2 |
| RGARCH-SSTD-RRV |  | 1.69 (0.19) | 4.28 (0.12) | **18.19 (0.01\*)** | 6 |  | 4.61 (0.03) | 6.99 (0.03) | **18.59 (0.01\*)** | 7 |
| RGARCH-GED-RRV |  | 0.78 (0.38) | 2.67 (0.26) | 13.81 (0.05) | 1 |  | 2.53 (0.11) | 3.64 (0.16) | 8.87 (0.26) | 4 |
| RGARCH-NIG-RRV |  | 1.69 (0.19) | 4.28 (0.12) | **18.1 (0.01\*)** | 5 |  | 3.51 (0.06) | 4.3 (0.12) | 12.35 (0.09) | 6 |
|  |  |  |  |  | 0.68 |  |  |  |  | 0.47 |
| WPD-DCS |  | **6.06 (0.01\*)** | **6.74 (0.03\*)** | **15.52 (0.03\*)** | 5 |  | 0.37 (0.54) | 1.06 (0.59) | 3.21 (0.86) | 1 |
| RGARCH-SSTD-RV |  | 1.05 (0.31) | 3.95 (0.14) | **20.33 (0\*\*)** | 1 |  | 2.6 (0.11) | 2.86 (0.24) | **17.27 (0.02\*)** | 3 |
| RGARCH-GED-RV |  | 0.64 (0.42) | 2.21 (0.33) | **16.54 (0.02\*)** | 3 |  | 1.66 (0.2) | 2.1 (0.35) | **17.28 (0.02\*)** | 5 |
| RGARCH-NIG-RV | 0.92 | 0.64 (0.42) | 2.21 (0.33) | **16.53 (0.02\*)** | 4 | 0.97 | 1.66 (0.2) | 2.1 (0.35) | **17.32 (0.02\*)** | 2 |
| RGARCH-SSTD-RRV |  | 1.55 (0.21) | 3.91 (0.14) | **15.07 (0.04\*)** | 7 |  | **4.99 (0.03\*)** | **6.46 (0.04\*)** | **16.48 (0.02\*)** | 6 |
| RGARCH-GED-RRV |  | 1.05 (0.31) | 3.95 (0.14) | **16.29 (0.02\*)** | 2 |  | 3.71 (0.05) | 3.84 (0.15) | **19.69 (0.01\*)** | 4 |
| RGARCH-NIG-RRV |  | 1.55 (0.21) | 3.91 (0.14) | **14.96 (0.04\*)** | 6 |  | 3.71 (0.05) | 3.84 (0.15) | **18.25 (0.01\*)** | 7 |
|  |  |  |  |  | 0.19 |  |  |  |  | 0.32 |
| WPD-DCS |  | 3.67 (0.06) | 4.35 (0.11) | 13.55 (0.06) | 5 |  | 1.71 (0.19) | 2.25 (0.32) | 1.69 (0.98) | 1 |
| RGARCH-SSTD-RV |  | 0.91 (0.34) | 1.71 (0.43) | 7.24 (0.4) | 1 |  | **4.22 (0.04\*)** | 4.89 (0.09) | **28.18 (0\*\*)** | 3 |
| RGARCH-GED-RV |  | 0.51 (0.47) | 1.62 (0.45) | 7.9 (0.34) | 3 |  | **4.22 (0.04\*)** | 4.89 (0.09) | **28.35 (0\*\*)** | 5 |
| RGARCH-NIG-RV | 0.93 | 0.51 (0.47) | 1.62 (0.45) | 7.93 (0.34) | 4 | 0.98 | 2.85 (0.09) | 3.54 (0.17) | **22.11 (0\*\*)** | 2 |
| RGARCH-SSTD-RRV |  | 1.41 (0.24) | 5.59 (0.06) | **24.4 (0\*\*)** | 7 |  | **5.8 (0.02\*)** | **6.24 (0.04\*)** | **33.02 (0\*\*)** | 6 |
| RGARCH-GED-RRV |  | 2 (0.16) | 5.52 (0.06) | **22.02 (0\*\*)** | 2 |  | **5.8 (0.02\*)** | **6.24 (0.04\*)** | **31.27 (0\*\*)** | 4 |
| RGARCH-NIG-RRV |  | 1.41 (0.24) | 5.59 (0.06) | **24.29 (0\*\*)** | 6 |  | **5.8 (0.02\*)** | **6.24 (0.04\*)** | **32.79 (0\*\*)** | 7 |
|  |  |  |  |  | 0.26 |  |  |  |  | 0.24 |

| Model | | | | | | | | | | |
|---|---|---|---|---|---|---|---|---|---|---|
| WPD-DCS | | 1.75 (0.19) | 2.42 (0.3) | 11.87 (0.1) | 5 | | **8.01 (0\*\*)** | **8.55 (0.01\*)** | 2.06 (0.96) | 1 |
| RGARCH-SSTD-RV | | 0.77 (0.38) | 2.66 (0.26) | 7.52 (0.38) | 1 | | **8.01 (0\*\*)** | **8.55 (0.01\*)** | **54.91 (0\*\*\*)** | 4 |
| RGARCH-GED-RV | | 0.77 (0.38) | 2.66 (0.26) | 7.66 (0.36) | 4 | | **8.01 (0\*\*)** | **8.55 (0.01\*)** | **55.09 (0\*\*)** | 5 |
| RGARCH-NIG-RV | 0.94 | 0.77 (0.38) | 2.66 (0.26) | 7.69 (0.36) | 3 | 0.99 | 5.72 (0.02) | 6.14 (0.05) | **20.47 (0\*\*)** | 2 |
| RGARCH-SSTD-RRV | | 1.89 (0.17) | 4.94 (0.08) | **19.42 (0.01\*)** | 7 | | **10.55 (0\*\*)** | **11.25 (0\*\*)** | **55.7 (0\*\*\*)** | 7 |
| RGARCH-GED-RRV | | 1.27 (0.26) | 2.74 (0.25) | 11.24 (0.13) | 2 | | **8.01 (0\*\*)** | **8.55 (0.01\*)** | **54.74 (0\*\*\*)** | 3 |
| RGARCH-NIG-RRV | | 1.89 (0.17) | 4.94 (0.08) | **19.22 (0.01\*)** | 6 | | **8.01 (0\*\*)** | **8.55 (0.01\*)** | **55.79 (0\*\*\*)** | 6 |
| | | | | | 0.37 | | | | | 0.63 |

Note: *, **, and *** represent statistical significance levels of 5%, 1%, and .1%, respectively. The rank tells the superiority of these four models under a default level ($\alpha = 0.15$) set in R function, and p-value helps to prove the non-rejection of this superiority. The p-value of the relevant test is indicated by the value in parentheses. Bold text indicates that it fails the test.

Table B8. Back testing results of out-of-sample daily VaR forecasts, SZ399001.

| Model | Alpha | LRuc statistics | LRcc statistics | DQ statistics | MCS | Alpha | LRuc statistics | LRcc statistics | DQ statistics | MCS |
|---|---|---|---|---|---|---|---|---|---|---|
| WPD-DCS | 0.9 | 2.76 (0.1) | **8.38 (0.02\*)** | **20.93 (0\*\*)** | 3 | 0.95 | 0.27 (0.61) | 4.56 (0.1) | **31.03 (0\*\*)** | 2 |
| RGARCH-SSTD-RV | | 1.34 (0.25) | **6.66 (0.04\*)** | **15.05 (0.04\*)** | 5 | | **4.44 (0.04\*)** | **7.49 (0.02\*)** | 13.56 (0.06) | 5 |
| RGARCH-GED-RV | | 1.34 (0.25) | **6.66 (0.04\*)** | **15.1 (0.03\*)** | 1 | | 3.44 (0.06) | **7.12 (0.03\*)** | 13.05 (0.07) | 1 |
| RGARCH-NIG-RV | | 1.34 (0.25) | **6.66 (0.04\*)** | **15.07 (0.04\*)** | 2 | | 3.44 (0.06) | **7.12 (0.03\*)** | 13.04 (0.07) | 3 |
| RGARCH-SSTD-RRV | | 2.42 (0.12) | **8.83 (0.01\*)** | **20.95 (0\*\*)** | 6 | | **9.49 (0\*\*)** | **21.08 (0\*\*)** | **54.58 (0\*\*\*)** | 7 |
| RGARCH-GED-RRV | | 2.42 (0.12) | **8.83 (0.01\*)** | **20.97 (0\*\*)** | 7 | | **9.49 (0\*\*)** | **21.08 (0\*\*)** | **54.77 (0\*\*\*)** | 6 |
| RGARCH-NIG-RRV | | 3.07 (0.08) | **8.61 (0.01\*)** | **21.63 (0\*\*)** | 4 | | **6.76 (0.01\*)** | **13.78 (0\*\*)** | **26.86 (0\*\*)** | 4 |
| | | | | | 0.67 | | | | | 0.38 |
| WPD-DCS | 0.91 | 1.95 (0.16) | 5.19 (0.07) | **19.76 (0.01\*)** | 2 | 0.96 | 1.7 (0.19) | 5.99 (0.05) | **40.02 (0\*\*)** | 1 |
| RGARCH-SSTD-RV | | 1.69 (0.19) | **8.76 (0.01\*)** | **21.08 (0\*\*)** | 5 | | 2.53 (0.11) | 6.1 (0.05) | **15.4 (0.03\*)** | 5 |
| RGARCH-GED-RV | | 1.69 (0.19) | **8.76 (0.01\*)** | **21.06 (0\*\*)** | 1 | | 2.53 (0.11) | 6.1 (0.05) | **15.32 (0.03\*)** | 3 |
| RGARCH-NIG-RV | | 1.69 (0.19) | **8.76 (0.01\*)** | **21.07 (0\*\*)** | 4 | | 2.53 (0.11) | 6.1 (0.05) | **15.42 (0.03\*)** | 2 |
| RGARCH-SSTD-RRV | | 3.67 (0.06) | **11.01 (0\*\*)** | **27.03 (0\*\*)** | 6 | | **7.2 (0.01\*)** | **13.89 (0\*\*)** | **30.67 (0\*\*)** | 7 |
| RGARCH-GED-RRV | | 3.67 (0.06) | **11.01 (0\*\*)** | **27.08 (0\*\*)** | 7 | | **7.2 (0.01\*)** | **13.89 (0\*\*)** | **31.23 (0\*\*)** | 6 |
| RGARCH-NIG-RRV | | 3.67 (0.06) | **11.01 (0\*\*)** | **26.92 (0\*\*)** | 3 | | **8.67 (0\*\*)** | **14.44 (0\*\*)** | **30.55 (0\*\*)** | 4 |
| | | | | | 0.77 | | | | | 0.27 |
| WPD-DCS | 0.92 | 1.23 (0.27) | 4.8 (0.09) | **19.31 (0.01\*)** | 1 | 0.97 | 3.71 (0.05) | 5.62 (0.06) | **45.23 (0\*\*)** | 1 |
| RGARCH-SSTD-RV | | 3.57 (0.06) | **10.64 (0\*\*)** | **25.52 (0\*\*)** | 5 | | **4.99 (0.03\*)** | **6.46 (0.04\*)** | 14.04 (0.05) | 5 |
| RGARCH-GED-RV | | 3.57 (0.06) | **10.64 (0\*\*)** | **25.5 (0\*\*)** | 4 | | **4.99 (0.03\*)** | **6.46 (0.04\*)** | 14 (0.05) | 3 |
| RGARCH-NIG-RV | | 3.57 (0.06) | **10.64 (0\*\*)** | **25.51 (0\*\*)** | 3 | | **4.99 (0.03\*)** | **6.46 (0.04\*)** | 14.05 (0.05) | 2 |
| RGARCH-SSTD-RRV | | **4.41 (0.04\*)** | **13.29 (0\*\*)** | **32.42 (0\*\*)** | 6 | | **7.99 (0\*\*)** | **14.05 (0\*\*)** | **34.29 (0\*\*)** | 7 |
| RGARCH-GED-RRV | | 3.57 (0.06) | **10.64 (0\*\*)** | **25.44 (0\*\*)** | 7 | | **9.69 (0\*\*)** | **18.5 (0\*\*)** | **50.47 (0\*\*\*)** | 6 |
| RGARCH-NIG-RRV | | 3.57 (0.06) | **10.64 (0\*\*)** | **25.45 (0\*\*)** | 2 | | **6.42 (0.01\*)** | **9.98 (0.01\*)** | **25.24 (0\*\*)** | 4 |
| | | | | | 0.54 | | | | | 0.17 |
| WPD-DCS | 0.93 | 0.28 (0.59) | 3.85 (0.15) | **22.34 (0\*\*)** | 2 | 0.98 | **4.22 (0.04\*)** | **7.94 (0.02\*)** | **33.03 (0\*\*)** | 1 |
| RGARCH-SSTD-RV | | 2 (0.16) | 5.52 (0.06) | 13.45 (0.06) | 6 | | **7.57 (0.01\*)** | **8.82 (0.01\*)** | **25.85 (0\*\*)** | 5 |
| RGARCH-GED-RV | | 1.41 (0.24) | 5.59 (0.06) | 13.15 (0.07) | 1 | | **7.57 (0.01\*)** | **8.82 (0.01\*)** | **25.56 (0\*\*)** | 3 |

| | | | | | | | | | |
|---|---|---|---|---|---|---|---|---|---|
| RGARCH-NIG-RV | | 1.41 (0.24) | 5.59 (0.06) | 10.84 (0.15) | 4 | **5.8 (0.02*)** | **6.84 (0.03*)** | **25.52 (0**)** | 2 |
| RGARCH-SSTD-RRV | | **5.33 (0.02*)** | **13.4 (0**)** | **33.42 (0**)** | 5 | **11.62 (0**)** | **13.1 (0**)** | **29.71 (0**)** | 7 |
| RGARCH-GED-RRV | | **5.33 (0.02*)** | **13.4 (0**)** | **33.44 (0**)** | 7 | **11.62 (0**)** | **13.1 (0**)** | **30.1 (0**)** | 6 |
| RGARCH-NIG-RRV | | **5.33 (0.02*)** | **13.4 (0**)** | **33.45 (0**)** | 3 | **9.51 (0**)** | **11.42 (0**)** | **28.4 (0**)** | 4 |
| | | | | | 0.52 | | | | 0.17 |
| WPD-DCS | 0.94 | 0.03 (0.86) | 1.51 (0.47) | **16.94 (0.02*)** | 2 | 0.99 | **4.9 (0.03*)** | 4.9 (0.09) | 2.42 (0.93) | 1 |
| RGARCH-SSTD-RV | | 2.61 (0.11) | **7.55 (0.02*)** | **14.57 (0.04*)** | 6 | **10.55 (0**)** | **11.25 (0**)** | **29.36 (0**)** | 5 |
| RGARCH-GED-RV | | 1.89 (0.17) | **7.65 (0.02*)** | **14.52 (0.04*)** | 1 | **10.55 (0**)** | **11.25 (0**)** | **28.9 (0**)** | 3 |
| RGARCH-NIG-RV | | 2.61 (0.11) | **7.55 (0.02*)** | **14.58 (0.04*)** | 3 | **8.01 (0**)** | **8.55 (0.01*)** | **27.31 (0**)** | 2 |
| RGARCH-SSTD-RRV | | **7.72 (0.01*)** | **16.87 (0**)** | **44.53 (0***)** | 5 | **13.33 (0**)** | **14.19 (0**)** | **63.01 (0**)** | 7 |
| RGARCH-GED-RRV | | **7.72 (0.01*)** | **16.87 (0**)** | **44.62 (0***)** | 7 | **19.49 (0**)** | **19.76 (0**)** | **74.62 (0**)** | 6 |
| RGARCH-NIG-RRV | | **7.72 (0.01*)** | **16.87 (0**)** | **44.49 (0***)** | 4 | **10.55 (0**)** | **11.25 (0**)** | **59.67 (0**)** | 4 |
| | | | | | 0.45 | | | | 0.23 |

Note: *, **, and *** represent statistical significance levels of 5%, 1%, and .1%, respectively. The rank tells the superiority of these four models under a default level ($\alpha = 0.15$) set in R function, and p-value helps to prove the non-rejection of this superiority. The p-value of the relevant test is indicated by the value in parentheses. Bold text indicates that it fails the test.

Table B9. P-values of LM test for out-of-sample parameter's scores of 30-min return.

| | SH000001-lambda | SZ399001-lambda |
|---|---|---|
| 1 | 1 | 0.28 |
| 2 | 1 | 1 |
| 3 | 0.69 | 0.41 |
| 4 | 1 | 0.7 |
| 5 | 0.34 | 0.97 |
| 6 | 0.93 | 0.96 |
| 7 | 0.69 | 0.31 |
| 8 | 0.95 | 0.57 |

Note: a p-value greater than 0.05 indicates that the null hypothesis that the sequence is not self-correlated is not rejected.